\documentstyle{ar}
\input epsf
\newcommand{\rstev}{\mbox{$\rs = \T{1800}$}}
\newcommand{\rssps}{\mbox{$\rs = \Ge{630}$}}

\newcommand{\ET}{\mbox{$E_{T}$} }

\newcommand{\AEA}{\mbox{$|\eta|$}}
\newcommand{\JETRAD}{\mbox{\sc JETRAD }}
\newcommand{\EKS}{\mbox{\sc EKS }}
\newcommand{\Ge}[1]{\mbox{#1 GeV}}
\newcommand{\T}[1]{\mbox{#1 TeV}}
\def\D0{D\O}
\def\ETmiss{{\rm {\mbox{$E\kern-0.57em\raise0.19ex\hbox{/}_{T}$}}}}

\newcommand{\ipb}{\mbox{$pb^{-1}$}}

\newcommand{\rs}{\mbox{$\sqrt{\rm s}$}}

\begin{document}

\title{Inclusive Jet and Dijet Production at the Tevatron}

\date{today}

\markboth{Gerald C. Blazey and Brenna L. Flaugher}
{Inclusive Jet and Dijet Production at the Tevatron}

\author{Gerald C. Blazey\affiliation{Department of Physics, Northern
Illinois University, DeKalb, Illinois, 60119} Brenna L.
Flaugher\affiliation{Fermi National Accelerator Laboratory,
Batavia, Illinois, 60115}}


\begin{abstract}
High energy jet distributions measured since 1992 at the Fermilab
Tevatron proton--antiproton collider are presented and compared to
theoretical predictions. The statistical uncertainties on these
measurements are significantly reduced relative to previous
results. The systematic uncertainties are comparable in size to the
uncertainty in the theoretical predictions. Although some
discrepancies between theory and measurements are noted, the
inclusive jet and dijet cross sections can be described by
quantum chromodynamics. Prospects for reducing the uncertainty in
the theoretical predictions by incorporating Tevatron measurements
into the proton parton distributions are discussed. Dijet
distributions, in excellent agreement with quantum chromodynamics,
set a 2.5 TeV limit on the mass of quark constituents.
\end{abstract}

\maketitle

\section{INTRODUCTION}

In the past seven years, quantum chromodynamics (QCD), the accepted
theory of quark and gluon interactions, has been confronted with a
set of precise and varied measurements of jet production from the
Fermi National Accelerator Laboratory proton--antiproton collider.
In $p\bar{p}$ collisions, jet production can be understood as a
point--like collision of a quark or gluon from the proton and a
quark or gluon from the antiproton. After colliding, these
scattered partons manifest themselves as sprays of particles or
``jets''. The extremely high energy of these interactions provides
an excellent opportunity to test our understanding of perturbative
QCD (pQCD). In particular, differences between QCD calculations and
measured jet processes reveal information about both the partonic
content of the proton and the nature of the strong interaction.
Further, unexpected deviations may signal the existence of new
particles or interactions down to distance scales of $10^{-19}$
meters or less.

Historically, jet measurements have involved tests of QCD at the
highest energies and searches for new physics. Prior to 1992, the
inclusive jet cross section, one of the most important hadron
collider results, had been measured at the ISR $pp$ collider at a
center--of--mass energy of $\rs = 63$ GeV \cite{AFS}, the CERN
$p\bar{p}$ collider at 546 and 630 GeV \cite{UA2, UA1}, and the Fermilab
Tevatron $p\bar{p}$ collider at 1800 GeV ~\cite{CDFINC89,
CDFINC92}. These data span a factor of twenty in beam energy and a
factor of two hundred in jet energy transverse to the proton beam
($E_T$) and, in general, are reasonably well described by QCD.

Studies of the correlations between the leading two jets of an
interaction also test QCD and provide opportunities to search for
new physics. Measurements of dijet distributions prior to 1992,
such as the invariant mass of the leading two jets or the angular
separation between the jets, are also in good agreement with QCD
\cite{UA1DIJET, UA2DIJET, CDFDJM, CDFANGULAR89, CDFANGULAR92}.
Using the model of Eichten, Lane, and Peskin ~\cite{COMP} these
data are used to search for composite quarks ~\cite{UA2, UA1,
CDFINC89, CDFINC92, CDFDJM, CDFANGULAR89, CDFANGULAR92}, with the
best limits coming from an inclusive jet cross section
measurement~\cite{CDFINC92}. In this model excess jet production at
large \ET relative to QCD is interpreted as the product of quark
constituent scattering. Because these data sets are relatively
small all searches are limited by statistical precision rather than
systematic uncertainties at the highest $E_T$. A summary of these
inclusive jet and dijet results can be found in the review of QCD
by Huth and Mangano ~\cite{HUTH}.

In 1992, with the advent of high luminosity data
collection at the Tevatron, a new era of precise $p\bar{p}$ jet
physics began. From 1992--1993 (Run 1A)
each of the two collider experiments
accumulated $\approx$ 20
$pb^{-1}$ of data and from 1994--1996 (Run 1B) each
accumulated $\approx$ 100 $pb^{-1}$. Both
data sets were taken at $\rs =$ 1800 GeV. At the end of Run 1B
a small portion of data ( $\approx$ 600 $nb^{-1}$) was
taken at $\rs =$ 630 GeV.
The two runs together represent a data set $\approx$ 50 times
larger than any preceding data set.

In contrast to previous tests of QCD,
the Run 1 jet measurements have statistical
uncertainties significantly smaller than experimental or
theoretical systematic uncertainties. In fact, the inclusive
jet analysis from the Run 1A sample yielded a significant
discrepancy between data and contemporaneous theoretical
predictions as well as good agreement with previous measurements
\cite{CDF1996}.  In addition to stimulating great excitement over
the possibility of departures from QCD, the result motivated a
re--evaluation of the uncertainties associated with inclusive jet
cross section calculations~\cite{SOPER96, STIRLING97, BERTRAM}.
Later measurements of the inclusive jet cross
section using the Run 1B samples stimulated further discussions
since one confirmed~\cite{CDF-preprints} the
Run 1A measurement while the other was well described by QCD.
In addition, subsequent studies have
indicated that QCD can describe the observed Run 1A cross section
through adjustments to the parton distribution functions (PDFs),
which represent the fraction of proton momentum carried by the
constituent quarks and gluons.

Independent of the actual high $E_T$ behavior of the inclusive jet
cross section, the Run 1A and 1B results have revealed large
uncertainties in the theoretical predictions. In particular, the
PDFs are derived from fits to data from many different experiments,
all of which are collected at low energy and extrapolated to
Tevatron jet energies. Representation of the uncertainties in the
resulting PDFs is a complex issue and has never been precisely
resolved. Consequently, the Tevatron collaborations have taken an
interest in derivation of the PDFs from information collected at
the Tevatron. Measurements of dijet mass distributions and dijet
cross sections at a variety of scattering angles provide
information on the momentum distributions of the partons. By
comparing the different measurements, constraints on the PDFs can
be established, and sensitivity to the presence of new physics
should ultimately be improved.

This review opens with a description of the theoretical framework
behind perturbative QCD and a discussion of the uncertainties
associated with predictions. Descriptions of the two detectors in
which these measurements were performed and jet reconstruction
algorithms are also provided. The review then devotes a section
each to the Run 1 measurements of the inclusive jet cross section,
the ratio of inclusive cross sections at different beam energies,
dijet differential cross sections, and dijet mass and angular
distributions. As will be seen, these measurements offer a detailed
look into the composition of the proton and the nature of the
strong interaction. A conclusion summarizes the results and
suggests future avenues of research.

\section{PERTURBATIVE QUANTUM CHROMODYNAMICS}

\subsection{Leading Order and Next--to--Leading Order
QCD}

\begin{figure}
\begin{center}
  \epsfxsize=5.0in
  \epsfysize=2.5in
  \epsffile{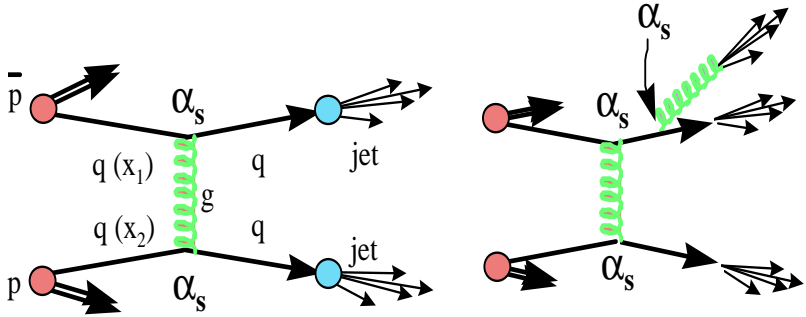}
  \caption{Factorization of the scattering process:  In this example
   incoming quarks with momentum fraction $x_1$ and $x_2$ of the incident
   hadrons scatter through gluon exchange and fragment into final
   state jets.}
\label{ORDER}
\end{center}
\end{figure}

The proton--antiproton interaction, a fairly general scattering
process, nicely introduces the concepts of leading order and
next--to--leading order pQCD.  As
illustrated in Fig.~\ref{ORDER}, inelastic scattering between a
proton and an antiproton can be described as an elastic collision
between a single proton constituent and single antiproton
constituent. These constituents are collectively referred to as
partons and in QCD are quarks and gluons. The non--colliding
constituents of the incoming proton and anti--proton are called
beam fragments or spectators.  Predictions for jet production are
given by folding experimentally determined parton distribution
functions $f$ with perturbatively calculated
two--body scattering cross
sections $\hat{\sigma}$ . See, for example, reference~\cite{kellisqcd}
for a detailed discussion.
The two ingredients can be formally
combined to calculate any cross section of interest:\
\begin{center}
$\sigma = \sum_{i,j} \int dx_{1}dx_{2}
f_{i}(x_{1},\mu^{2}_{F})f_{j}(x_{2},\mu^{2}_{F})
\hat{\sigma} [ x_{1}P, x_{2}P, \alpha_{s}(\mu^{2}_{R}),
Q^{2}/\mu^{2}_{F},Q^{2}/\mu^{2}_{R}]$.
\end{center}
The parton distribution functions $f_{k}(x,\mu^{2}_{F})$ describe
the momentum fraction $x$ of the incident hadron momentum $P$
carried by a parton of type $k$ (gluons or quarks). The PDF is
defined in terms of the factorization scale $\mu_{F}$. The hard
two--body cross section is a function of the momentum carried by
each of the incident partons $xP$, the strong coupling parameter
$\alpha_{s}$, the scale $Q$ characterizing the energy of the hard
interaction, and the renormalization scale $\mu_{R}$. Final state
partons manifest themselves as collimated streams or ``jets'' of
particles. This formal description includes no explicit
hadronization or fragmentation functions to describe the transition
from partons to jets.  For most high $E_T$ measurements these
effects are small and jets are identified as partons.

The two--body scattering has been illustrated in Fig.~\ref{ORDER}
with a leading order (LO) graph.
The cross section for this process is
proportional to two powers of the strong coupling
parameter, $\alpha_{s}$, which come from the two vertices.
Although useful, the LO picture
is too simple and has large normalization uncertainties.  Next--to--leading
order (NLO) or O$(\alpha^{3}_{s})$ calculations include one additional
parton emission.
A schematic example is shown in the second illustration
of Fig.~\ref{ORDER}. Here a final state quark has radiated an
additional gluon and the entire scattering process is proportional
to $\alpha^{3}_{s}$.
Depending on the proximity of the other partons, a ``jet" could
result from one or two (combined) partons. This in turn results in
parton level predictions for the shape of jets, and for the
effects of clustering parameters.

A complete theoretical prediction of jet production should not
depend on internal calculational details; however, this is not the
case for fixed order perturbative QCD, which depends on the choice
of factorization and renormalization scales. The factorization
scale, a free parameter, determines how the contributions of
initial state radiation are factorized between the PDFs and
$\hat{\sigma}$. The renormalization scale is also arbitrary and is
related to choices in the theoretical calculations designed to
control or renormalize ultraviolet singularities.
Typically, as in this report, $\mu_{R}$ and $\mu_{F}$ are
set equal to each other, $\mu_{R} = \mu_{F} = \mu $, and $\mu$ is chosen to
be of the same order as $Q$.  The sensitivity of the theoretical
predictions to $\mu$ is often taken as a measure of
the uncertainty from the contributions of higher order terms.

Predictions for jet production at NLO have been derived by Ellis,
Kunszt, and Soper \cite{EKS}; Aversa {\it et al.} \cite{AVERSA};
and Giele, Glover, and Kosower \cite{GGK}. The NLO predictions are
much less sensitive to the choice of $\mu$~\cite{EKS}. The ``{\sc
EKS}'' program by Ellis, Kunszt, and Soper~\cite{SOPERPAGE} can
generate analytic predictions for jet cross sections as a function
of final state parameters.  The ``{\sc JETRAD}'' program by Giele,
Glover, and Kosower~\cite{JETRADPAGE} generates weighted ``events''
with final state partons. Cross sections are calculated by
generating a large number of events as a function of final state
parameters. All predictions in this document have been generated
either with \EKS or {\sc JETRAD}. The two programs agree within a
few percent\cite{BERTRAM}.

\subsection{Theoretical Choices}

A NLO QCD calculation requires selection of several input
parameters including specification of a parton clustering
algorithm, a perturbative scale, and PDFs. Taken together these
choices can result in approximately 30\% variations in the
theoretical predictions.  The single largest uncertainty is due to
the PDFs. These input choices and the resultant changes in the
predictions are discussed in the following paragraphs. The
variations are illustrated by comparing predictions for the
inclusive jet cross section. As in the experimental results to
follow, the cross section is reported as a function of jet \ET and
pseudorapidity $\eta=-{\rm ln}({\rm tan}(\theta/2))$ where $\theta$
is the angle between the jet and the proton beam, and is summed
over $\phi$, the azimuthal angle around the beam. For reference,
$\eta=0$ for jet production at 90 degrees relative to the proton
beam.

\subsubsection{PARTON CLUSTERING}

The NLO predictions may include two or three final state partons.
To convert the partons to the equivalent of jets measured in a
detector, a clustering algorithm is employed. The Snowmass cone
algorithm~\cite{SNOWMASS} was proposed as a way to minimize the
difference between theoretical predictions and experimental
measurements.  In theory calculations, two partons which fall
within a cone of radius R in $\eta$--$\phi$ space (R =
$\sqrt{\Delta\eta^2 + \Delta\phi^2}$ and $\Delta\eta$ and
$\Delta\phi$ are the separation of the partons in pseudorapidity
and azimuthal angle) are combined into a ``jet".  A standard cone
radius of $R=$ 0.7 is used by both collider experiments for most
measurements. A consequence of this algorithm is that partons have
to be at least a distance of 2R apart to be considered as separate
jets.

\begin{figure}
\centerline{
  \epsfxsize=4.0in
  \epsfysize=2.5in
  \epsffile{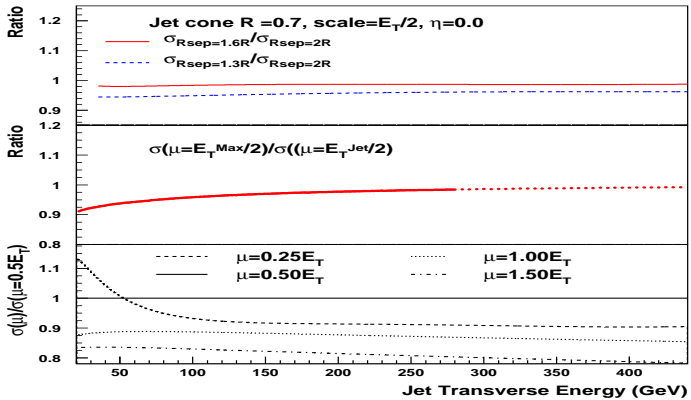}}
  \caption{ NLO theoretical predictions for the inclusive jet cross section.
  Each prediction is normalized relative to the NLO \EKS calculation with
  CTEQ4M,  $\mu=\ET^{jet}/2$, $R_{sep}=1.3$.}
\label{THEORY_ERR}
\end{figure}

Subsequent studies indicated that the experimental clustering
algorithms (described in later sections) were more efficient at
separating nearby jets~\cite{RSEP, D0SHAPE} than the idealized
Snowmass algorithm. In other words,
two jets would be identified even though they were separated
by less than 2R. An additional parameter, $R_{sep}$, was introduced
in the QCD predictions to mimic the experimental
effects of cluster merging and
separation. Partons within $R_{sep}
\times R$ were merged into a jet, while partons separated by more
than $R_{sep}
\times R$ were identified as two individual jets. As described in
detail in the references, a value of $R_{sep}$ = 1.3 was found to
give the best agreement with the data for cross section and jet
shape measurements \cite{RSEP}. This corresponds in the data to the
50\% efficiency point in jet separation~\cite{D0SHAPE}, i.e. two
jets within $R_{sep}$=1.3 of one another are merged 50\% of the
time and identified as two individual jets 50\% of the time. At
separations of 1.0R the algorithms nearly always merge the two jets
and at values of 1.6R they nearly always identify two separate jets.
The top panel of Figure~\ref{THEORY_ERR} shows the change in the prediction
for the inclusive jet cross section using the EKS program when
$R_{sep}$ is decreased from 2 to 1.3.  The result is primarily a
normalization change of 5 to 7\%.

\subsubsection{THE SCALE}

Because a NLO calculation is truncated at order $\alpha_{s}^{3}$
there is some residual dependence on the scale $\mu$ at which the
calculation is performed. The scale is usually taken to be
proportional to $\ET^{jet}$ (or just $E_T$), maximum transverse jet
energy $\ET^{max}$ in a given event, or the total center--of--mass
energy. Other choices for the scale are possible. To study the
scale dependence of the predictions, the magnitude of the scale is
varied by a multiplicative coefficient, common choices are $\mu
= 0.5\ET^{jet}$ and $0.5\ET^{max}$. The middle panel of
Fig.~\ref{THEORY_ERR} shows a mild \ET dependence at the 2--9\%
level on the definition of the scale. The lowest panel shows that
typical variations of the multiplicative coefficient lead to
5--20\% shifts in the cross section with only small changes in
shape above 100 GeV.

\subsubsection{PARTON DISTRIBUTION FUNCTIONS}

Parton distribution functions are derived from global fits
to data primarily from
deep inelastic scattering (DIS) fixed target and
electron--proton collider experiments.  In DIS, a lepton probe is
used to sample the partonic structure of the target hadrons.
For use at the Tevatron, the resulting distributions must
be evolved from the
low energy DIS results to the high energy range of
jet measurements.

As the data from DIS experiments increases and improves, new parton
distribution functions are derived. Some recent PDFs also
incorporate data from the Tevatron. The result is a plethora of
PDFs, each with its own specific list of data included in the fit,
assumptions as to the value of $\alpha_s$, functional forms for the
quark and gluon momentum distributions, and assumptions concerning
the contributions of the gluons, which are not constrained by the
DIS results. The most recent PDFs include the most precise data and
combined knowledge, and supplant the previous PDFs.

Several groups have analyzed the available
data and produced families of candidate PDFs. Within a family,
the individual PDFs represent the range of predictions resulting from
changes in one of the input assumptions such as the value of $\alpha_s$.
Additional variations come from differences in
the input data sets or the relative weighting between
the data sets. The CTEQ2M \cite{CTEQ2M},
CTEQ3M \cite{CTEQ3M} and MRSA$^{\prime}$
\cite{MRSAP} families of PDFs incorporate data published before
1994 and do not include Tevatron jet data.  The CTEQ4M PDFs
\cite{CTEQ4M} include data published before 1996; CTEQ4HJ
\cite{CTEQ4M} additionally has a high $x$ gluon adjustment designed
to accommodate the Run 1A high \ET jet cross section measurements.
The MRST \cite{MRST} PDF family utilizes data published before 1998
and adds a contribution for a putative initial transverse momentum
of the partons.

\begin{figure}
\centerline{
  \epsfxsize=3.5in
  \epsfysize=2.5in
  \epsffile{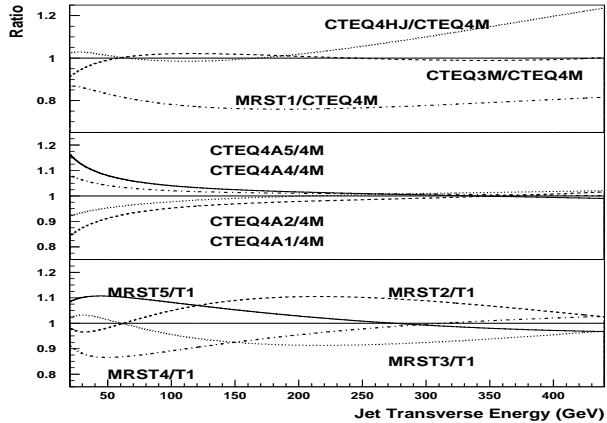}}
  \caption{ NLO theoretical predictions for the inclusive jet cross section
  using different PDFs.
  Each prediction is normalized relative to the NLO \EKS calculation with
  CTEQ4M.  All used  $\mu=\ET/2$ and $R_{sep}=1.3$}
\label{PDF_ERR}
\end{figure}

Comparison between observed cross sections and NLO predictions with
alternate PDFs provide some insight into the quark and gluon
composition of the proton. As shown in the Fig.~\ref{PDF_ERR} the
PDFs can result in $\approx$ 20\% variations in jet cross sections;
typical variations within a set are on
the order of 5--10\%. Although the quark
distributions were thought to be well known and the contribution of
the gluons was expected to be small, investigations~\cite{CTEQ98,
YANG98} have shown that there are uncertainties at high $x$, which
were ignored in the derivation of early PDFs.  In addition,
studies\cite{CTEQ4M} revealed that the gluon distribution could be
adjusted to give a significant increase in the jet cross section at
high $E_{T}$, while maintaining reasonable agreement with the low
energy data sets.

\section{DETECTORS AND JET DEFINITIONS}

\subsection{The CDF and D0 Detectors}

The Collider Detector at Fermilab (CDF) \cite{CDF} and the D0
Detector \cite{D0} at the Fermi National Accelerator Laboratory
Tevatron $p\bar{p}$ collider are complementary general purpose
detectors designed to study a broad range of particle physics topics.
The Tevatron is designed with six locations at which the
counter--rotating proton and antiproton beams can collide.
The CDF and D0 detectors each occupy one of these
interaction regions;
D0 takes its name from the alphanumeric designation of its
interaction region.
Each detector is
comprised of a series of concentric sub--systems, which surround
the $p\bar{p}$ interaction region.
Immediately outside the Tevatron beam pipe each
detector has a tracking system designed to detect charged
particles. The tracking systems are surrounded by calorimeters, which
measure the energy and direction of electrons, photons, and jets.
Finally, the calorimeters are covered with muon tracking systems.

Luminosity monitors at each detector measure the beam exposure.
These monitors are located near the beam lines and detect particles
from the $p\bar{p}$ collisions.  The luminosity at a given site is
given by dividing the luminosity monitor event rate by that portion
of the total $p\bar{p}$ cross section to which the luminosity
monitors have acceptance.  A direct comparison of jet cross
sections at the two experiments is complicated by the fact that the
collaborations adopt slightly different values for the total
$p\bar{p}$ cross section~\cite{ALBROW,CDFLUM,D0LUM}. As a result,
the CDF jet cross section measurements are 2.7\% higher than the
corresponding D0 jet cross section measurements~\cite{ALBROW}.

In the following
sections we mention primarily the central tracking systems and the
calorimeters since these systems are the most important
for jet identification and reconstruction. As will be seen, CDF has
a high resolution particle tracking system, which makes a crucial
contribution to jet energy calibration. On the other hand, and in a
complementary fashion, D0 has highly segmented, uniform, and thick
calorimetry well suited to {\it in--situ} jet energy measurements.

Both calorimeters are segmented into projective towers. Each tower
points back to the center of the nominal interaction region and is
identified by its pseudorapidity and azimuth. The polar angle
$\theta$ in spherical coordinates is measured from the proton beam
axis, and the azimuthal angle $\phi$ from the plane of the
Tevatron. The towers are further segmented longitudinally into
individual readout cells. The energy of a calorimeter tower is
obtained by summing the energy in all the cells of the same
pseudorapidity and azimuth.  The transverse energy or $E_T$ of a
tower is the sum of the \ET components of the cells. These
components are calculated by assigning each cell of a tower a
massless four--vector with magnitude equal to the energy deposited
and with the direction defined by the unit vector pointing from the
event origin to the center of the tower segment.

\subsubsection{D0 in Run 1}

The central tracking volume ($\AEA \approx 2$) of D0 includes the
vertex chamber, transition radiation detector, and central drift
chamber, arranged in three cylinders concentric with the beamline,
and two forward drift chambers.  The non--magnetic tracker is
compact, with an outer radius of 75 cm and an overall length of 270
cm centered on $z=0$.  Without the need to measure momenta of
charged particles, the prime considerations for tracking were good
two track resolving power, high efficiency, and good ionization
energy measurement.  For jet physics the central tracker is used to
find event vertices.

\begin{figure}
\centerline{
  \epsfxsize=3.0in
  \epsfysize=2.5in
  \epsffile{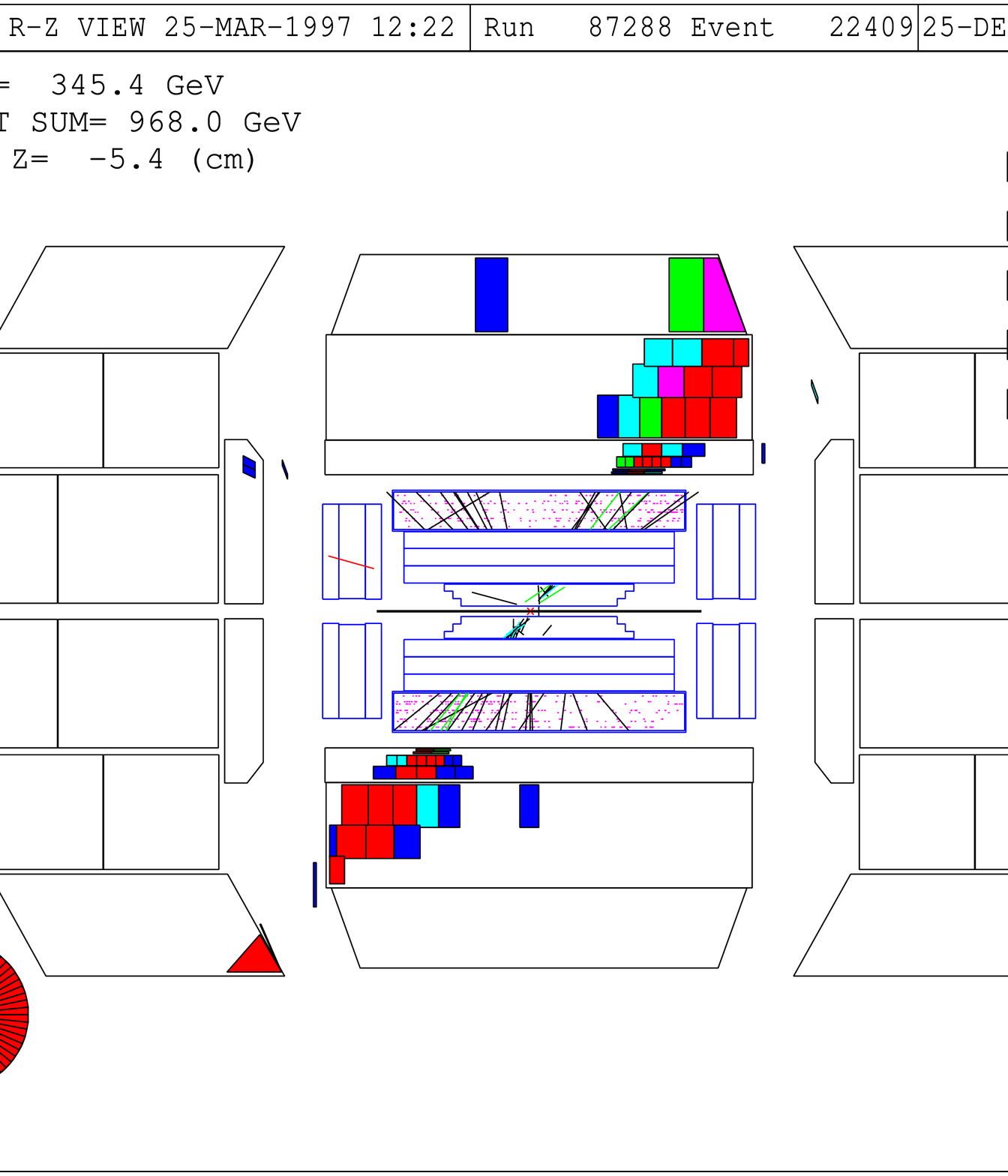}}
  \caption{The distribution of energy in a dijet event in the D0 detector.}
\label{TWOJET}
\end{figure}

Jet detection in the D0 detector primarily utilizes liquid
argon--uranium \\ calorimeters which are  hermetic, finely
segmented, thick, uniform and have unit gain. These sampling
calorimeters are composed of alternating layers of liquid argon and
absorber. The particles in a jet interact with the absorber, and
the resulting particle shower ionizes the liquid argon. This
ionization is detected as a measure of the jet energy. The
calorimeter is enclosed in three cryostats: the central calorimeter
(CC) covers $\AEA \leq 1.2$, and the end calorimeters (ECs) extend
the coverage to $\AEA \leq 4.4$. The calorimeters have complete
$2\pi$ azimuthal coverage. Between the cryostats the calorimeter
sampling is augmented by scintillator tiles with segmentation
matching the argon calorimeters.  Fig.~\ref{TWOJET} is a graphical
representation of the calorimeter modules described below.

The CC includes three concentric rings of calorimeter modules.
There are 32 electromagnetic calorimeter modules (CCEM) in the
inner ring, 16 fine hadronic modules (CCFH) in the middle ring, and
16 coarse hadronic modules (CCCH) in the outer ring. The CCEM and
CCFH calorimeters have uranium absorber plates, and the CCCH has
copper absorber plates. Longitudinal segmentation includes seven
samples: four in the CCEM, three in the CCFH, and one in the CCCH.
At $\eta=0$ the CC has a thickness of 7.2 nuclear absorption
lengths.  The calorimeter cells are segmented into $\Delta\eta
\times \Delta\phi = 0.1 \times 0.1$ except at shower maximum in the
third layer of the CCEM where the segmentation is $0.05 \times
0.05$. The calorimeter segmentation is designed to form projective
towers of size $\Delta\eta \times \Delta\phi = 0.1 \times 0.1$
geometry which point to $z=0$.  In the CC because of the finer
resolution in the third layer of the CCEM each tower is comprised
of eleven cells.

The two mirror--image ECs contain four module types.  An
electromagnetic module (ECEM) surrounding the beam pipe is backed
by an inner hadronic module (ECIH) which also surrounds the pipe.
Outside the ECEM and ECIH are concentric rings of 16 middle and
outer hadronic modules (ECMH and ECOH).  The ECEM and ECIH have
uranium absorber plates and four longitudinal segments.  The ECMH
has four uranium absorber segments nearest the interaction point
and one stainless steel absorber segment behind.  The ECOH employs
stainless steel plates and includes three longitudinal segments. At
\AEA = 2.0 a jet encounters 8--9 longitudinal segments and a
nuclear interaction thickness of $\approx 10 \lambda_{o}$.  The
transverse segmentation is similar to that of the CC.  The CC(EC)
electron energy resolution is $14.8(15.7)\%/ \sqrt{E}$ plus
0.3(0.3)\% added in quadrature and for hadrons $47.0(44.6)\%/
\sqrt{E}$ plus 4.5(3.9)\% added in quadrature.

The calorimeter response to the different types of particles is the
most important aspect of the D0 jet energy calibration.
Electromagnetically interacting particles, like photons ($\gamma$)
and electrons deposit most of their energy in the electromagnetic
sections of the calorimeters.  Hadrons, by contrast, lose energy
primarily through nuclear interactions and extend over the full
interaction length of the calorimeter. In general, the calorimeter
response to the electromagnetic ($e$) and nuclear or hadronic
components ($h$) of hadron showers is not the same.
Non--compensating calorimeters have a response ratio $e/h$ greater
than one and suffer from non-Gaussian event--to--event fluctuations
in the fraction of energy lost through electromagnetic production.
Such calorimeters give a non-Gaussian signal distribution for
hadrons and jets.  The D0 calorimeter is nearly compensating and
the hadronic and jet response are well described by a Gaussian
distribution~\cite{D0ENERGY}. In fact, single jet resolution as
measured with dijet events has a Gaussian line--shape and is
approximately 7\% at 100 GeV and 5\% at 300 GeV
~\cite{D0RESOLUTION}.

The D0 jet calibration requires correction for the hadronic
response of the jet; showering of energy outside the cone; and
subtraction of an offset which can be attributed to instrumental
effects, pile--up from previous beam--beam crossings, additional
interactions, and spectator energy. These corrections are derived
primarily {\it in--situ}~\cite{D0ENERGY}. The correction for
hadronic response begins with the electromagnetic calibration of
the calorimeter which is performed with dielectron and diphoton
decays of the $Z$ and $\pi^{0}$ resonances since electrons and
photons deposit their energy in the electromagnetic calorimeters.
The hadronic response for centrally produced $\gamma$--jet events
is derived from data using \ET balance between the photon and jet.
Because of the rapidly falling $\gamma$--jet cross section the
central calorimeter balance technique is limited to jet energies
below 150 GeV. Balancing central, well measured photons against
high rapidity jets permits the energy calibration to exceed 300
GeV. The response calibration is extended above 300 GeV using
simulated $\gamma$--jets events. For a R=0.7, 100 GeV jet at
$\eta=0$ the hadronic response is about $0.85 \pm 0.01$. The
showering correction for this jet is about $1 \pm 1$\% and is
derived from a study of jet profiles. The total offset correction
for a R=0.7 jet at $\eta=0$ as determined from a study of data sets
with various triggering and luminosity conditions is about $2.6 \pm
0.3$ GeV. The contribution from the underlying event or spectator
energy alone is roughly $0.9 \pm 0.1$ GeV.  At $\eta=0$ the mean
total energy correction for a R=0.7, 100 GeV jet is $15.0 \pm
1.7$\% and decreases slowly with energy.

\subsubsection{CDF in Run 1}

The CDF central ($\AEA \approx 1$) tracking system has three
sub--systems located within a 1.5 T magnetic field, that is provided by a
superconducting solenoid coaxial with the beam.  Nearest the beam,
a four--layer silicon microstrip vertex (SVX)
detector~\cite{SVX-det} occupies the radial region between 3.0 and
7.9 cm from the beamline and provides precision $r-\phi$ measurements.
Outside the SVX a vertex drift chamber (VTX) provides
$r$--$z$ tracking information and is used to locate the position of
the $p\bar{p}$ interaction (event vertex) in $z$, along the beam
line. Both the SVX and the VTX are mounted inside a 3.2 m long
drift chamber called the central tracking chamber (CTC). The radial
coverage of the CTC is from 31 to 132 cm. The momentum
resolution~\cite{CDF} of the SVX--CTC system is $\delta P_{T}/
P_{T}^2 = [(0.0009P_{T})^{2} + (0.0066)^{2}]^{1/2}$ where $P_{T}$
has units of GeV/c. The CTC provides {\it in--situ} measurement of
the calibration and response of the calorimeter to low energy
particles (where test beam information is not available) along with
measurements of jet fragmentation properties.

The solenoid and tracking volumes of CDF are surrounded by
calorimeters, which cover $2\pi$ in azimuth and $\AEA \leq 4.2$.
The central electromagnetic (CEM) calorimeter covers $\AEA \leq
1.1$ and is followed at larger radius by the central hadronic
calorimeters (CHA and WHA) which cover $\AEA \leq 1.3$. These
calorimeters use scintillator as the active medium.  The CEM
absorber is lead and the CHA/WHA absorber is iron. The calorimeters
are segmented into units of 15 degrees in azimuth and 0.1
pseudorapidity. Two phototubes bracket each tower in $\phi$ and the
average of the energy in the two tubes is used to determine the
$\phi$ position of energy deposited in a tower. The interaction
length of both the CHA and WHA is $4.5\lambda_{o}$. Electron energy
resolution in the CEM is $13.7\% / \sqrt{E}$ plus 2\% added in
quadrature. For hadrons the single particle resolution depends on
angle and varies from roughly $50\% / \sqrt{E}$ plus 3\% added in
quadrature in the CHA to $75\% / \sqrt{E}$ plus 4\% added in
quadrature in the WHA. In the forward regions ($1.1<\AEA<4.2 $)
calorimetric coverage is provided by gas proportional chambers. The
plug electromagnetic (PEM) and hadronic calorimeters (PHA) cover
the region from $1.1<\AEA<2.4$. The forward electromagnetic (FEM)
and hadronic calorimeters (FHA) cover the region from
$2.2<\AEA<4.2$. The segmentation of these detectors is roughly 0.1
in $\eta$ and 5 degrees in $\phi$.

\begin{figure}
\centerline{
  \epsfxsize=4.0in
  \epsfysize=4.0in
  \epsffile{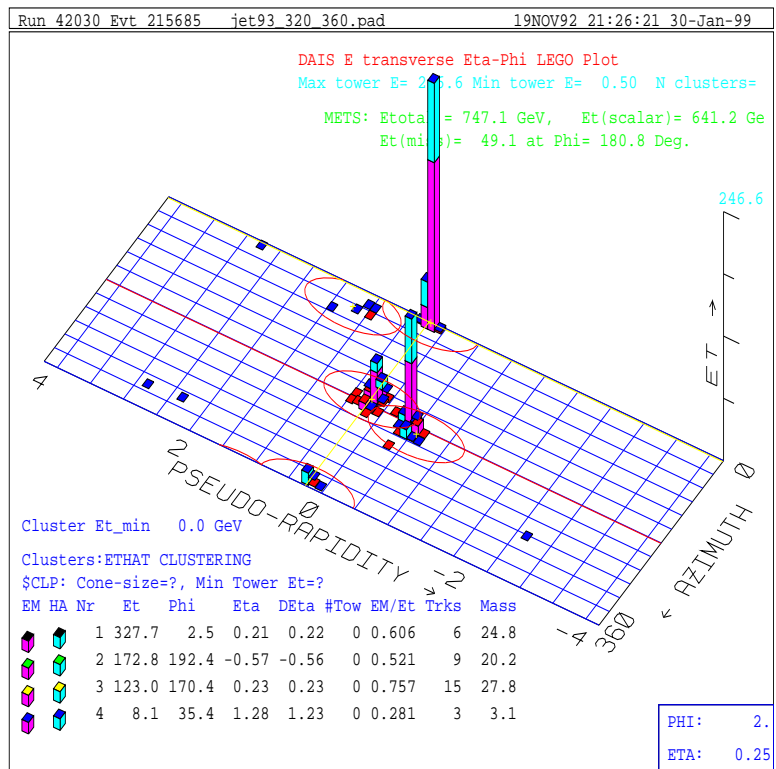}}
  \caption{A 3--jet event in the CDF detector.  A jet clustering cone
of radius 0.7 is shown around each jet.}
\label{CDFLEGO}
\end{figure}

Figure~\ref{CDFLEGO} shows a
3--jet event in the CDF calorimeter. In this ``lego'' plot the
calorimeter is ``rolled out'' onto the $\eta$--$\phi$ plane.  The tower
height is proportional to the $E_T$ deposited in the tower. The
darker and lighter shading of each tower corresponds to the $E_T$
of the electromagnetic and hadronic cells of the tower
respectively.
The oval around each clump of energy indicates the jet
clustering cone described in the next section.

Observed jet energies are corrected for a number of effects
including the calibration and response of the detectors and the
background energy from the remnants of the $p\bar{p}$ interaction.
In the central detectors corrections associated with detector
response are obtained from a Monte Carlo program, which is tuned to
give good agreement with the data from electron and hadron test
beams ($E_{beam}$ ranged from 10 to 227 GeV), and to data from an
{\it in--situ} study using CTC track momenta and isolated  hadrons
with 400 MeV/c$<P_T<$ 10 GeV/c. In addition, the Monte Carlo jet
fragmentation parameters were tuned to agree with both the number
and momenta of particles observed in jets. The resulting simulation
was then used to determine ``response functions'', which represent
the response of the calorimeter to jets as a function of jet $E_T$.
The width of the response functions represents the jet energy
resolution and can be expressed as $\sigma$ = 0.1$E_T$ + 1
GeV~\cite{CDFINC92}. Calibration of the plug and forward detectors
is achieved through a dijet balancing technique~\cite{cdf-4jet}.
The $E_T$ of jets in the plug and forward detectors is balanced
against jets in the central region whose calibration is pinned by
the tracking information. The energy left in a jet cone from the
hard interaction spectators is called the ``underlying event
energy". A precise theoretical definition of this quantity does not
exist. Experimentally, it has been estimated from events in which
there is no hard scattering, and from the energy in cones
perpendicular to the jet axis in dijet events. For a cone of radius
0.7 the contribution to the jet $E_T$ is of order 1 GeV. An
uncertainty of 30\% is assigned to this quantity to cover
reasonable variations in the definition of this quantity. The
magnitude of the total jet energy corrections and the corresponding
uncertainty depend on the $E_T$ of the jets and their location in
the detector, and on the specific analysis. Typically, the jet \ET
correction factor is in the range of 1.0 to 1.2.

\subsubsection{CDF and D0 in Run 2}

The Fermilab Tevatron is undergoing major improvements for Run 2, which is
slated to begin in the year 2000. The energy of the proton and
antiproton beams will increase from 900 to 1000 GeV and the
instantaneous luminosity will increase by at least a factor of 5.
The expected Run 2 data sample is 2$fb^{-1}$ in the first 2 years
of operation.

Both the CDF and D0 collaborations are improving their detectors
for Run 2 operations. Among other things, CDF is replacing the gas
calorimeters with scintillating tiles and is closing gaps between
calorimeters.  CDF is also extending the tracking such that
calibration of the new plug detector can be accomplished using
tracks, as in the central detector.  D0 is also upgrading a number
of detector components. For jet measurements the most important
upgrade is the replacement of the nonmagnetic tracking system with
a magnetic tracker.  The tracker (including a high resolution
silicon strip detector, a scintillating fiber tracker, and
electron/photon preshower detectors) will enable calibration of the
calorimeter with single particles. Thus, in Run 2 the detectors
will be more similar than in Run 1: CDF will have more uniform
calorimetry and D0 will have magnetic tracking. The result of these
improvements should be an overall reduction of jet measurement
uncertainties.

\subsection{The Experimental and Theoretical Definitions of a Jet}

From the experimental point of view, jets are the manifestation of
partons as showers of electromagnetic and
hadronic matter.  Jets are observed as clusters of energy located
in adjacent detector towers. Typically, a jet contains tens of
neutral and charged pions (and to a lesser extent kaons), each of
which showers into multiple cells. A single jet illuminates roughly
20 towers (this corresponds to approximately 40 calorimeter cells
in the CDF detector and over one hundred detector cells in the D0
detector). Fig.~\ref{TWOJET} illustrates the energy distribution of
a two--jet event in the D0 detector. Each rectangular outline
represents the energy deposited in calorimeter cells at fixed
$\eta$ and depth. The eye quite naturally
clusters the energy into two jet--like objects.  However, for the
purposes of jet cross section measurements a more quantitative
definition of the jet is required.

For high $E_T$ measurements at CDF and D0, cone algorithms are used
to identify jets in the calorimeters. Cone algorithms operate on
objects in pseudorapidity and azimuth space, such as particles,
partons, calorimeter cells, or calorimeter towers.  For the sake of
simplicity towers will be used in the following discussions. The
CDF and D0 algorithms are both based on the Snowmass algorithm
\cite{SNOWMASS}. This algorithm defines a jet as those towers
within a cone of radius $\Delta R = \sqrt{\Delta \phi^{2} + \Delta
\eta^{2}}$ = 0.7. The jet \ET is the sum of the transverse energies
of the towers in the cone, and the location of the jet is defined
by the \ET weighted $\eta$ and $\phi$ centroids:
\begin{center}
\[E_T^{jet} = \sum_{i} \ET^{i}\]

\[\eta_{jet} = (\sum_{i} \ET^{i}\eta_{i}) / \ET^{jet}\]

\[\phi_{jet} = (\sum_{i}\ET^{i}\phi_{i}) / \ET^{jet}.\]
\end{center}
The sum over $i$ is over all towers that are within the jet radius
$R$.

In Figure~\ref{CDFLEGO}
the clustering cone is shown by the oval around each jet and
is centered on the cluster centroid.
The two overlapping cones in this event
indicate that the two nearby clusters have been identified as two separate
jets. The Snowmass algorithm doesn't specify cell thresholds or the
handling of such overlapping jet cones. These details must be dealt
with according to the needs of individual experiments, and led to
the introduction of the $R_{sep}$ parameter in the theoretical
calculations.

In the D0 experiment, jets are defined in two stages
~\cite{D0SHAPE}. In the first or clustering stage, all the energy
that belongs to a jet is accumulated, and in the second stage, the
$\eta,$ $\phi$ and \ET of the jet are defined.
The clustering consists of the following steps: 1)
Calorimeter towers with \ET $\geq$ 1 GeV are enumerated. Starting
with the highest \ET tower, preclusters are formed by adding
neighboring towers within a radius of R=0.3. 2) The jet direction
is calculated for each precluster using the sums defined for the
Snowmass algorithm. 3) All the energy in the towers in a cone of
radius R=0.7 around each precluster
is accumulated and used to recalculate $\eta$ and
$\phi$. 4) Steps 2 and 3 are repeated until the jet direction is
stable.  Finally, in the second stage, the jet energy and
directions are calculated according to the equations:\\
\begin{center}

\[E^{jet} = \sum_{i} E^{i},~~~\ET^{jet} = \sum_{i} E_T^{i}\]

\[\tan\theta_{jet} = \frac{\sqrt{ (\sum_{i}E^{i}_{x})^{2}
+(\sum_{i}E^{i}_{y})^{2} }} {\sum_{i}E^{i}_{z} } \]

\[\phi_{jet} = \tan^{-1} [ \sum_{i}E^{i}_{y}/ \sum_{i}E^{i}_{x}]^{2} \]

\end{center}
where $E^{i}_{x}=E_{i}\sin(\theta_{i})\cos(\phi_{i})$,
$E^{i}_{y}=E_{i}\sin(\theta_{i})\sin(\phi_{i})$, and
$E^{i}_{z}=E_{i}\cos(\theta_{i})$. Studies have shown that for
$\eta \leq \sim 1$ the final D0 jet directions are, within
experimental errors, equal to the Snowmass directions. Overlapping
jets are merged if more than 50\% of the smaller jet \ET is
contained in the overlap region.  If less than 50\% of the energy
is contained in the overlap region, the jets are split into two
jets and the energy in the overlap region is assigned to the
nearest jet.  After merging or splitting the jet directions are
recalculated.

The CDF cluster algorithm \cite{CDFCONE} has two stages of jet
identification similar to the D0 algorithm.  The first stage
consists of the following steps: 1) a list of towers with $E_T>$1.0
GeV is created; 2) preclusters are formed from an unbroken chain of
contiguous seed towers with continuously decreasing tower $E_T$; if
a tower is outside a window of $7\times7$ towers surrounding the
seed it is used to form a new precluster; 3) the preclusters are
grown into clusters by finding the \ET weighted centroid and
collecting the energy from all towers with more than 100 MeV within
R=0.7 of the centroid; 4) a new centroid is calculated from the set
of towers within the cone and a new cone drawn about this position;
steps 3 and 4 are repeated until the set of towers contributing to
the jet remains unchanged; 5) overlapping jets are merged if they
share $\geq$75\% of the smaller jet's energy;  if they share less,
the towers in the overlap region are assigned to the nearest jet.
In the second stage, the final jet parameters are computed.  The
angles and energy are calculated as in the D0 algorithm, however
the jet \ET is given by:
\[ \ET = E\sin\theta_{jet}.\]
One philosophical
difference between the CDF and Snowmass algorithms is that
the CDF jets have mass ($E_T \neq P_T$).
Studies~\cite{SNOWMASS} found that
the CDF clustering algorithm and the Snowmass algorithm
were numerically very similar.

Both the CDF and D0 algorithms are based on the Snowmass algorithm,
but the reality of jets of hadrons measured by a finitely segmented
calorimeter necessitates the introduction of additional steps and
cuts.  Apart from small definitional details, the most significant
difference between the Snowmass algorithm and its experimental
implementations is the handling of cluster merging and separation.
To simulate these effects in the NLO calculations the $R_{sep}$
parameter was introduced.  Considering the complexity of the
hadronic showers is it remarkable that a NLO calculation, with only
2 or 3 partons in the final state and a single additional parameter
can provide a good description of the observed jet shapes and cross
sections~\cite{RSEP}.

\section{THE INCLUSIVE JET CROSS SECTION AT 1800 GEV}

\subsection{Introduction}

The inclusive jet cross section represents one of the
most basic tests of QCD at a hadron--hadron collider.
It reflects the probability of observing a hadronic jet of a given
\ET and rapidity in a $p\bar{p}$ collision.   The term
inclusive indicates that all the jets in an event are included in
the cross section measurement and that the presence of additional
non--jet objects (for example electrons or muons) is irrelevant.
Theoretical calculations are normally expressed in terms of the
invariant cross section: \
\begin{center}
$E d^{3}\sigma / dp^{3}$
\end{center}
where $E$ and $p$ are the jet energy and momentum. In terms of
the natural experimental variables, the cross section is given by
\begin{center}
$d^{2}\sigma / d\ET d\eta$
\end{center} which is related to the first expression by,
\begin{center}
$E d^{3}\sigma / dp^{3} = (1/2 \pi E_{T})d^{2}\sigma / d\ET d\eta.$
\end{center}
Massless jets have been assumed ($P_{T}$ = $E_T$);
the $2\pi$ comes from the integration
over the azimuthal angle $\phi$.
The measured cross section is simply the
number of jets $N$ observed in an $\eta$ and
\ET interval normalized by the total luminosity exposure, $\cal{L}$:
\begin{center}
$ d^2\sigma / d\ET d\eta = N / (\Delta\ET
\Delta\eta \cal{L})$.
\end{center}

The CDF and D0 inclusive jet analyses place similar requirements on the
events and jets selected for calculation of the cross sections.
Both collaborations eliminate poorly measured events by requiring
the event vertices to be near the center of the detector.
Backgrounds (such as cosmic rays) are eliminated by rejecting
events with large missing $E_{T}$.  Spurious jets from cosmic ray
or instrumental backgrounds are eliminated with quality cuts based
on jet shapes. Corrections are made to the measured cross sections
to account for the event and jet detection inefficiencies,
mismeasurement of the jet energies, and for energy falling in a jet
cone from other sources (such as remnants of the $p\bar{p}$
collision, or additional $p\bar{p}$ interactions).  No corrections
are made for partons showering outside the jet cone as this should
be included in the NLO theoretical calculations.

We begin our discussion of the inclusive jet cross section with
measurements by the CDF collaboration.  The Run 1A measurement
stimulated interest due to a discrepancy with
QCD predictions at high $E_T$. A subsequent
measurement by the D0 collaboration with Run 1B data is well
described by theoretical predictions, while the preliminary CDF Run 1B
result is in good agreement with the 1A measurement.
In the next sections, the CDF and D0 measurements are each described
in some detail
and then comparisons of the two results are presented.

\subsection{The CDF Cross Section}

In 1996 CDF published the inclusive jet cross section measured from
the Run 1A data sample~\cite{CDF1996}
for jet $E_T$ from 15 to 440 GeV in the central
rapidity range 0.1$\leq |\eta| \leq$ 0.7. This analysis followed the
same procedures as previous measurements~\cite{CDFINC89, CDFINC92}
for correcting the cross section and for estimating the systematic
uncertainties. This process will be described briefly below.  More
detail is available in
references~\cite{CDFINC89,CDFINC92,CDF1996,CDF-XT-ANALYSIS}.

The measured jet $E_T$ spectrum requires corrections for energy
mismeasurement and for smearing effects caused by finite $E_T$
resolution.  This is accomplished with the ``unsmearing procedure"
described in \cite{CDF-XT-ANALYSIS}. As mentioned earlier, a Monte
Carlo simulation program~\cite{CDFINC92} tuned to the CDF data is
used to determine detector response functions. A trial true
(unsmeared) spectrum is smeared using these response functions and
compared to the raw data.  The parameters of the trial spectrum are
iterated to obtain the best match between the smeared trial
spectrum and the raw data. The corresponding unsmeared curve is
referred to as the ``standard curve'', and is used to correct the
measured spectrum.
The simultaneous correction for response and resolution
produces a result which is independent of the
$E_T$ binning used in the measurement while preserving the
statistical uncertainty on the measured cross section.
For jet $E_T$ between 50 and about 300 GeV these
corrections increase both the $E_T$ and the cross section by
$\approx$10\%.  At lower and higher $E_T$ the cross section
corrections are larger due to the steepening of the spectrum.

\begin{figure}
\centerline{
  \epsfxsize=3.5in
  \epsfysize=2.0in
  \epsffile{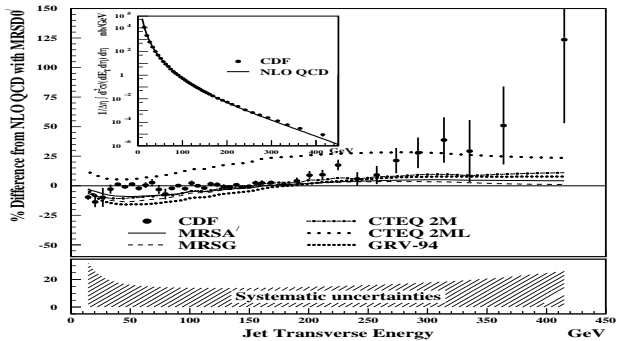}}
  \caption{Inclusive jet cross section measured by CDF using the Run 1A
data sample compared to predictions from the \EKS program with $\mu
= E_T^{jet}/2$, $R_{sep}$ = 2.0 and MRSD0'. The error bars
represent the statistical uncertainty on the cross section. The
shaded band represents the quadrature sum of the correlated
systematic uncertainties. Also shown is the percentage difference
between the predictions using other PDFs and the prediction with
MRSD0'.}
\label{CDFINC1A}
\end{figure}

Figure~\ref{CDFINC1A} shows the CDF measurement of the inclusive
jet cross section from the 20 \ipb~Run 1A data sample compared to a
NLO prediction from the EKS program with $\mu=E_T/2$, $R_{sep}$ =
2.0 and MRSD0' PDFs.  The inset shows the cross section on a
logarithmic scale.  There is good agreement between data and theory
over eleven orders of magnitude.  The main figure shows the
percentage difference between the data (points) and theory. The
bars on the points represent the statistical uncertainty. While
excellent agreement is observed below 200 GeV, an excess of events
over these theoretical predictions is observed at high $E_T$.
Predictions with other PDFs which were available at the time are
shown by the additional curves. For these predictions the
percentage from the default theory (MRSD0') is shown. As can be
seen from the figure, the best agreement below 200 GeV is with the
MRSD0' PDF. Other PDFs agree less well, and none show the rise at
high $E_T$ observed in the data.

\begin{figure}
\centerline{
  \epsfxsize=3.5in
  \epsfysize=2.5in
  \epsffile{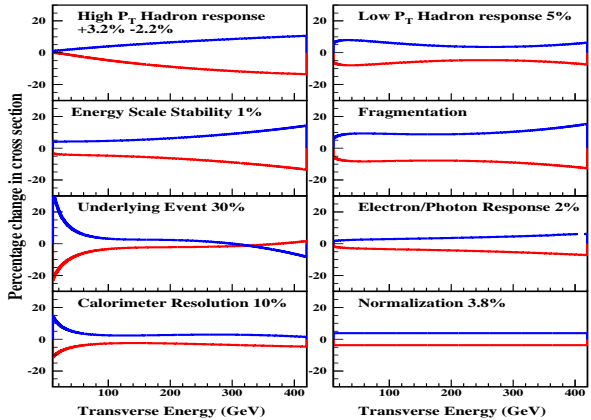}}
  \caption{The percentage change in the Run 1A inclusive jet cross section
when various sources of systematic uncertainty are changed by $\pm$1
standard deviation from their nominal value.}
\label{CDFINCSYS1A}
\end{figure}

The systematic uncertainties on the Run 1A cross section were
evaluated using the procedures in reference \cite{CDF-XT-ANALYSIS}.
In short, new parameter sets are derived for $\pm 1$ standard
deviation shifts in the unsmearing function for each source of
systematic uncertainty. The parameters for the Run 1A cross section
are given in Reference~\cite{CDF1996}.
Figures~\ref{CDFINCSYS1A}(a--h) show, for the Run 1A data sample,
the percentage change from the standard curve as a function of
$E_T$ for the seven largest systematic uncertainties:  (a) charged
hadron response at high $P_T$; (b) the calorimeter response to
low--$P_T$ hadrons; (c) $\pm 1\%$ on the jet energy for the
stability of the calibration of the calorimeter; (d) jet
fragmentation functions used in the simulation; (e) $\pm 30\%$ on
the underlying event energy in a jet cone; (f) detector response to
electrons and photons and (g) modeling of the detector jet energy
resolution. An eighth uncertainty, an overall normalization
uncertainty of $\pm$3.8\%, was derived from the uncertainty in the
luminosity measurement\cite{WZXSEC} ($\pm$3.5\%) and the efficiency
of the acceptance cuts ($\pm$1.5\%). The eight uncertainties arise
from different sources and are not correlated with each other.  The
1$\sigma$ shifts are evaluated by changing only one item at a time
in the Monte Carlo simulation, such as high $p_T$ hadron response.
The resulting uncertainty in the unsmeared cross section is thus
100\% correlated from bin to bin, but independent of the other
seven uncertainties.

To analyze the significance of the Run 1A result, the CDF
collaboration used four normalization--independent,
shape--dependent statistical tests~\cite{CDF1996}. The eight
sources of systematic uncertainty are treated individually to
include the $E_T$ dependence of each uncertainty. The effect of
finite binning and systematic uncertainties are modeled by a Monte
Carlo calculation. Between 40 and 160 GeV, the agreement between
data and theory is $>$80\% for all four tests.  Above 160 GeV,
however, each of the four methods yields a probability of 1\% that
the excess is due to a fluctuation. If the test is performed for
other PDFs, agreement at low $E_T$ is reduced, as is the
significance of the excess at high $E_T$. The best agreement at
high $E_T$ for the curves shown is with CTEQ2M\cite{CTEQ2M} which
gives 8\%, but the low $E_T$ agreement is reduced to 23\%.  The
excess of events observed at high $E_T$ initiated a re--evaluation
of the uncertainties in the PDFs, particularly at high $x$. One
outcome of this re--examination was CTEQ4HJ. This PDF
incorporates the same low energy DIS data as in CTEQ4M, but the
high \ET jet data is weighted to accentuate its contribution to the
global parton distribution fit in the high $Q$ and $x$ region.

\begin{figure}
\centerline{
  \epsfxsize=3.5in
  \epsfysize=2.5in
    \epsffile{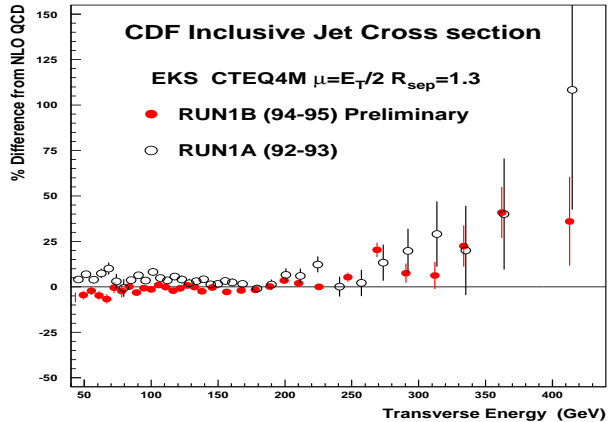}}
  \caption{The preliminary Run 1B inclusive jet cross
section compared to the published Run 1A data and to a QCD
prediction from the \EKS program with $\mu$= $E_T^{jet}/2$,
$R_{sep}$, = 1.3 and CTEQ4M.}
\label{CDFINC1B}
\end{figure}

\begin{figure}
\centerline{
  \epsfxsize=3.5in
  \epsfysize=2.5in
  \epsffile{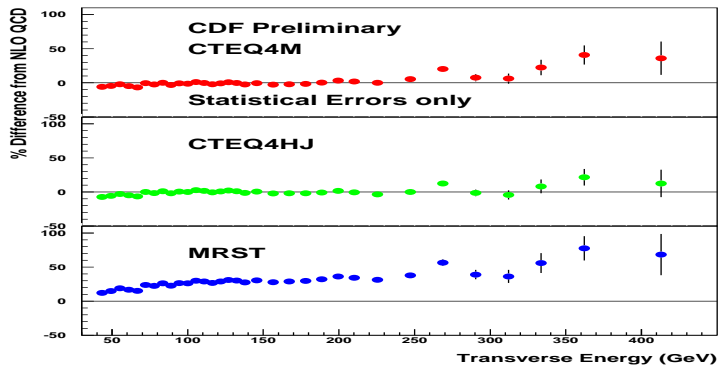}}
  \caption{The percent difference between the preliminary Run 1B
inclusive jet cross section and QCD predictions from the \EKS
program with $\mu$= $E_T^{jet}/2$, $R_{sep}$ = 1.3, and a variety
of current PDFs.}
\label{CDFINC1BPDF}
\end{figure}

Figure~\ref{CDFINC1B} shows the preliminary Run 1B
result~\cite{CDFNEW} compared to the Run 1A results and to CTEQ4M
(a more modern PDF than the ones in Figure~\ref{CDFINC1A}). The Run
1A and 1B data sets are in excellent agreement. Only statistical
uncertainties are shown. The greatly reduced statistical error bars
on the Run 1B data are due to the five--fold increase of luminosity
between the two runs. Analysis of the Run 1B data follows exactly
the sequence of the Run 1A data with some additional corrections
specific to the Run 1B running conditions. The Run 1B systematic
uncertainties are very similar to the uncertainties derived for the
Run 1A sample. Figure~\ref{CDFINC1BPDF} shows the Run 1B result
compared to predictions with other recent PDFs. Although
CTEQ4HJ is a bit contrived the good agreement with the data
demonstrates
the flexibility of the PDFs and the ability of the QCD calculations
to describe these Run 1A and Run 1B inclusive jet cross sections.
Quantitative comparisons between the Run 1B data and theoretical
predictions are underway.

\subsection{The D0 Cross Section}

In 1998 the D0 collaboration finalized a 92 \ipb~measurement of the
inclusive jet cross section~\cite{D01998}. The D0 analysis differs
from the CDF analysis in that the spectrum is corrected
independently for energy calibration and then for distortion by
finite jet energy resolution.  After passing the various jet and
event selection criteria each jet is corrected individually for the
average response of the calorimeter.  As mentioned in Section 3 the
jet response was determined with well measured photon--jet events.
However, the background free, energy corrected $\ET$ spectrum still
remains distorted by jet energy resolution.  The distortion is
corrected by first assuming that an ansatz function $(A \ET ^{-B})
\cdot (1 - 2\ET/\sqrt{s} )^{C}$ will describe the actual \ET
spectrum, then smearing it with the measured resolution, and
finally comparing the smeared result with the measured cross
section.  The procedure is repeated by varying parameters $A, B,$
and $C$ until the best fit is found between the observed cross
section and the smeared trial spectrum. At all $\ET$, the
resolution (measured by balancing $\ET$ in jet events) is well
described by a Gaussian distribution; at \Ge{100} the standard
deviation is 7 GeV. The ratio of the initial ansatz to the smeared
ansatz is used to correct the cross section on a bin--by--bin
basis~\cite{D0RESOLUTION}. The resolution correction reduces the
observed cross section by (13$\pm$3)\% [(8$\pm$2)\%] at \Ge{60}
[\Ge{400}].

\begin{figure}
\centerline{
\epsfysize=3.0in
\epsfxsize=4.0in
\leavevmode\epsffile{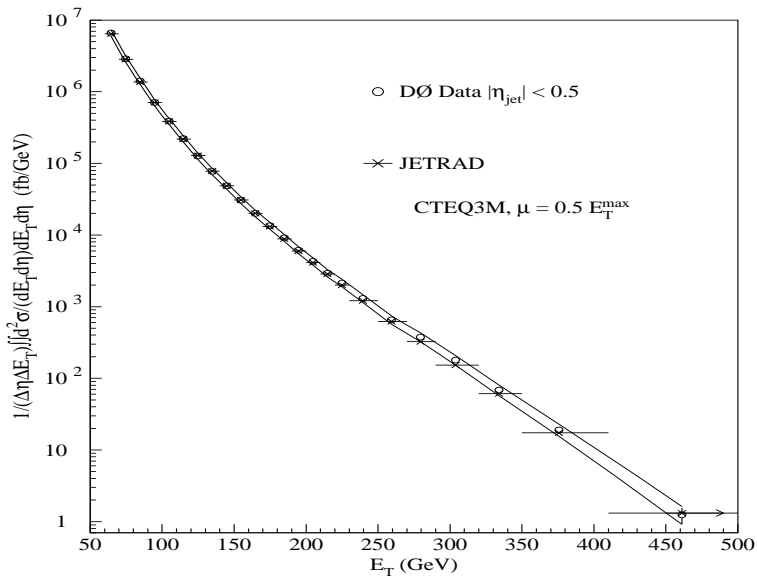}}
\caption[]{The D0 1800 GeV, $ \AEA \leq 0.5 $
    inclusive cross section.  Statistical uncertainties are invisible on this
    scale.  The solid curves represent the $\pm 1\sigma$ systematic
    uncertainty band on the data.}
\label{D0XSEC}
\end{figure}

Fig.~\ref{D0XSEC} shows the final inclusive jet cross section as
measured by the DO collaboration in the rapidity region $\AEA \leq
0.5$ \cite{D01998}.  This rapidity interval was chosen since the
detector is uniformly thick (seven or more interaction lengths with
no gaps) and both jet resolution and calibration are precise.  The
figure shows a theoretical prediction for the cross section from
\JETRAD. There is good agreement over seven orders of magnitude.
For the calculation shown here $\mu = 0.5 E_{T}^{\rm max}$,
the PDF is CTEQ3M, and $\cal{R}_{\rm{sep}}$=$1.3$.

\begin{figure}
\centerline{
\epsfysize=3.0in
\epsfxsize=4.0in
\leavevmode\epsffile{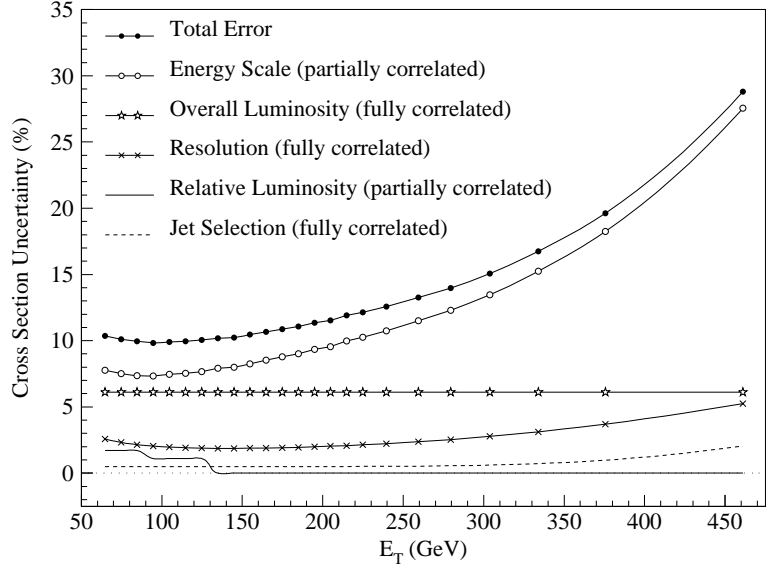}}
\caption[]{Contributions to the D0 , $ \AEA \leq 0.5 $ cross section
uncertainty plotted by component.}
\label{D0UNC}
\end{figure}

Fig.~\ref{D0UNC} shows the cross section uncertainties. Each curve
represents the average of nearly symmetric upper and lower
uncertainties. The energy scale uncertainty which varies from 8\%
at low $\ET$ to 22\% at 400 GeV dominates all other sources of
uncertainty, except at low $\ET$, where the 6.1\% luminosity
uncertainty is of comparable magnitude. The total systematic error
is 10\% at 100 GeV and 23\% at 400 GeV. Although the individual
errors are independent of one another each error is either 100\% or
nearly 100\% correlated point--to--point, and the overall
systematic uncertainty is highly correlated.
Table~\ref{TAB:D0CORRELS} shows that the bin--to--bin correlations
in the full uncertainty for representative $\ET$ bins are greater
than 40\% and positive.  The high degree of correlation will prove
a powerful constraint on data--theory comparisons.

\begin{table}
\begin{center}
\caption{D0 Inclusive jet cross section total uncertainty
correlations.}
\vspace{0.2cm}
\begin{tabular}{|c|c|c|c|c|c|}
\hline
\ET(GeV) & ~64.6 & 104.7 & 204.8 & 303.9 & 461.1 \\ \hline
~64.6 & 1.00  & 0.96  & 0.85  & 0.71  & 0.40  \\ \hline
104.7 & 0.96  & 1.00  & 0.92  & 0.79  & 0.46  \\ \hline
204.8 & 0.85  & 0.92  & 1.00  & 0.91  & 0.61  \\ \hline
303.9 & 0.71  & 0.79  & 0.91  & 1.00  & 0.67  \\ \hline
461.1 & 0.40  & 0.46  & 0.61  & 0.67  & 1.00  \\ \hline
\end{tabular}
\label{TAB:D0CORRELS}
\end{center}
\end{table}

\begin{figure}
\centerline{
\epsfysize=3.0in
\epsfxsize=4.0in
\leavevmode\epsffile{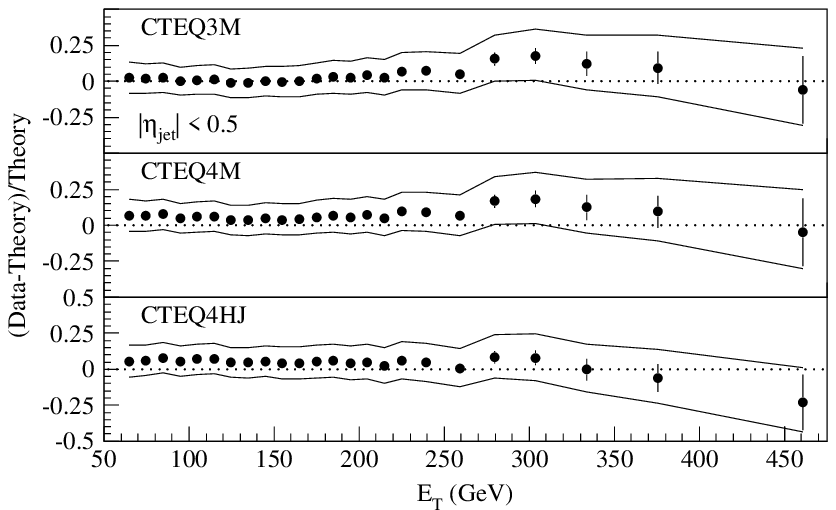}}
\caption[]{Difference between D0 data and \JETRAD QCD predictions
with $R_{sep}$ = 1.3 and $\mu = E_T^{max}/2$ normalized to
predictions for the range $|\eta|<$0.5. The bands represent the
total experimental uncertainty.}
\label{D0DMT}
\end{figure}

\begin{figure}
\centerline{
\epsfysize=3.0in
\epsfxsize=4.0in
\leavevmode\epsffile{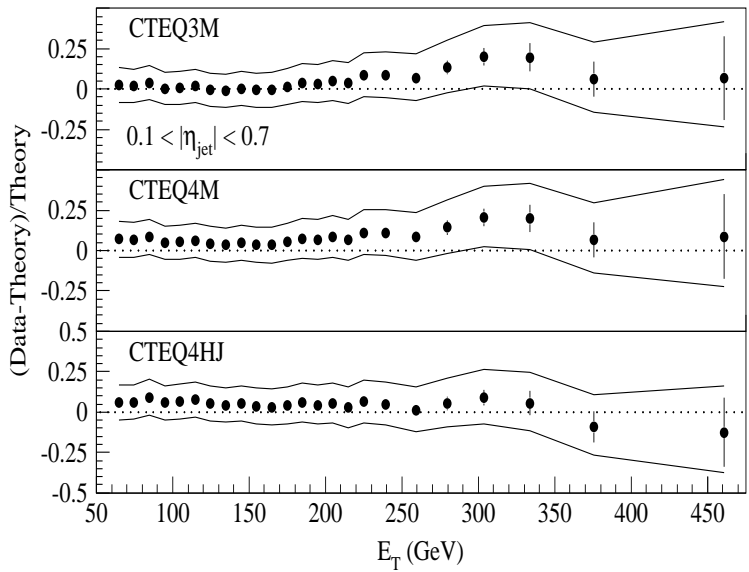}}
\caption[]{Difference between D0 data and \JETRAD QCD predictions
with $R_{sep}$ = 1.3 and $\mu = E_T^{max}/2$ normalized to
predictions for the range 0.1$<|\eta|<$0.7. The bands represent the
total experimental uncertainty.}
\label{D0DMT2}
\end{figure}

Figure \ref{D0DMT} shows the ratios $(D-T)/T$ for the D0 data ($D$)
and \JETRAD NLO theoretical predictions ($T$) based on the CTEQ3M,
CTEQ4M and CTEQ4HJ for the region $\AEA \leq 0.5 $. Figure
\ref{D0DMT2} shows the same ratios for the $0.1 \leq \AEA
\leq 0.7 $ data. Given the experimental and theoretical
uncertainties, the predictions are in agreement with the data; in
particular, the data above 350 GeV show no indication of a
discrepancy relative to QCD. The D0 collaboration has
quantitatively compared the data and theory with a $\chi^{2}$ test
incorporating the uncertainty covariance matrix. The matrix
elements are constructed by analyzing the mutual correlation of the
uncertainties in Fig.~\ref{D0UNC} at each pair of \ET values.
Table~\ref{TAB:D0CHI} lists $\chi^{2}$ values for several \JETRAD
predictions incorporating various PDFs. Each comparison has 24
degrees of freedom.  The \JETRAD predictions have been fit to a
smooth function of $E_T$.  All five predictions describe the $\AEA
\leq 0.5$ cross section very well (the probabilities for $\chi^{2}$
to exceed the listed values are between 47 and 90\%). A similar
measurement in the $0.1 \leq \AEA \leq 0.7$ interval is also well
described (probabilities between 24 and 72\%).  The probabilities
calculated by comparing the data to \EKS predictions for $\mu =
(0.25,0.5, 1.0) \times E_{T}^{\rm max}$ and $\mu = (0.25,0.5,1.0)
\times E_{T}^{\rm jet}$ are all greater than 57\%~. Perturbative
QCD is in good agreement with the data with or without large $x$
enhancements to the PDFs.

\begin{table}
\begin{center}
\caption{ $\chi^{2}$ comparisons between \JETRAD and
DO , $\AEA \leq 0.5 $ and $0.1 \leq \AEA \leq 0.7 $ data for $\mu
= 0.5E_{T}^{\rm max}$, $\cal{R}_{\rm{sep}}$=$1.3\cal{R}$, and
various PDFs.  There are 24 degrees of freedom. }
\label{TAB:D0CHI}
\vspace{0.2cm}
\begin{tabular}{|c|c|c|}
\hline
PDF     & $\AEA \leq 0.5 $ & $0.1 \leq \AEA \leq 0.7 $\\ \hline
CTEQ3M   & 23.9 & 28.4  \\ \hline CTEQ4M  & 17.6 & 23.3  \\ \hline
CTEQ4HJ & 15.7 & 20.5  \\ \hline MRSA\'  & 20.0 & 27.8  \\ \hline
MRST    & 17.0 & 19.5  \\ \hline
\end{tabular}
\end{center}
\end{table}
\vspace*{3pt}

\subsection{Comparison of the CDF and D0 Measurements}

The two inclusive jet analyses differ in an important but
complementary way. The CDF analysis utilizes a Monte--Carlo program
carefully tuned to collider and test beam data, to correct for
detector response.  In contrast, D0 corrects for calorimeter
response by direct utilization of collider data.  It is worth
mentioning that both collaborations invested years of time and
effort developing these procedures.  The CDF technique capitalizes
on the excellent tracking capabilities of the detector. The
tracking is used to directly measure the jet fragmentation
functions as well as verify test beam calibrations of the
calorimeter modules.  Additional checks of the Monte Carlo
simulation come from comparisons to collider jet $E_T$ balancing
data.  The uniformity of the D0 calorimetry permits a precise
collider data--based measurement of jet response and resolution at
all energies and rapidities. This uniformity permits transfer of
the jet energy calibration at low to moderate jet energies in the
central calorimeter to all pseudorapidity with a missing \ET
technique. The forward calorimetry provides an opportunity for
direct calibration of high energy jets relative to these central
photons.  The uniformity, linearity and depth of the detector also
assures that resolution functions are relatively narrow and
Gaussian.

\begin{figure}
\centerline{
\epsfysize=4.5in
\epsfxsize=4.5in
\leavevmode\epsffile{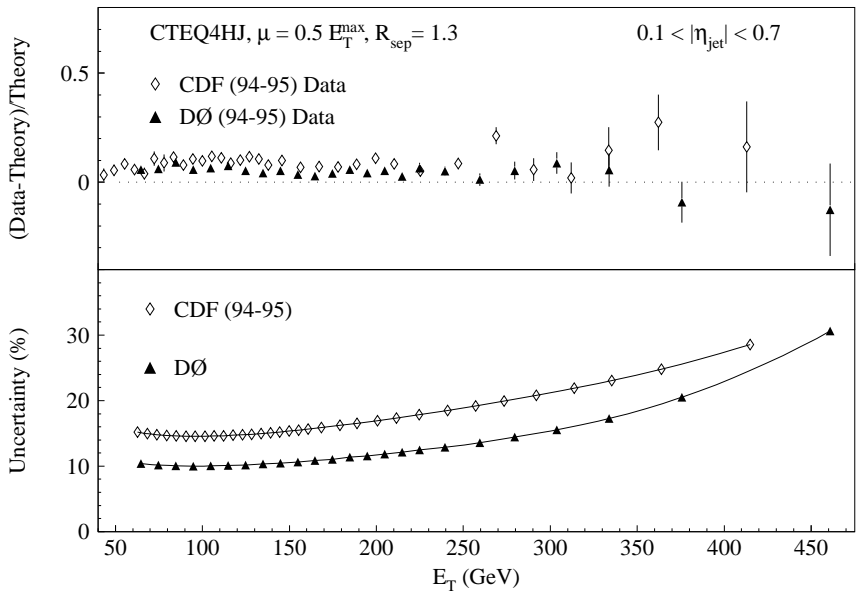}}
\vspace{-1.3in}
\caption[]{Top: comparisons of D0 and CDF data to \JETRAD in the
region 0.1$<|\eta|<$0.7.  Bottom: the quadrature sums of the D\O\
and CDF uncertainties.}
\label{D0CDF}
\end{figure}

The top half of Fig.~\ref{D0CDF} shows $(D-T)/T$ for the Run 1B D0
and CDF data sets in the $0.1 \leq \AEA \leq 0.7$ region relative
to a JETRAD calculation using CTEQ4HJ, $\mu=0.5 E_{T}^{\rm max}$,
and $\cal{R}_{\rm{sep}}$=$1.3$. Note that there is outstanding
agreement between the nominal values for \ET $\leq$ 350 GeV.  At
higher \ET the two results diverge but not significantly, given the
statistical and systematic errors. This impression is fortified by
a direct comparison of the magnitude of the D\O\ and CDF
uncertainties as shown in the second half of the figure. To
quantify the degree of agreement, the D0 collaboration has carried
out a $\chi^{2}$ comparison between their data and the nominal
curve describing the central values of the CDF 1B data. The nominal
curve is used instead of the data points because each of the
measurements report the cross section at different values of jet
$E_T$. Comparison of the D0 data to the nominal Run 1B curve, as
though it were theory yields a $\chi^2$ of 41.5 for 24 degrees of
freedom. A "statistical-error-only" comparison of the D0 and CDF
data is approximated by calculating the value of the CDF curve at
the D0 $E_T$ points and assuming the statistical uncertainty on the
CDF and D0 data are equivalent (the D0 statistical errors are
multiplied by $\sqrt{2}$.) When the 2.7\% relative normalization
difference~\cite{ALBROW} is removed, this statistical-error-only
$\chi^2$ is 35.1 for 24 degrees of freedom, a probability of 5.4\%.
When the systematic uncertainties in the covariance matrix are
expanded to include both the D0 and CDF uncertainties and the D0
statistical errors are increased by $\sqrt{2}$
the $\chi^{2}$ equals 13.1 corresponding to a probability
of $96$\%.

Considering the complexities of these analyses, the overall
agreement is remarkable!  Given the present flexibility of the PDFs
and the mixed agreement between the data and theory at the highest
$x$ and $Q$, no clear indication of a high \ET deviation with QCD
can be inferred. By using the high $E_T$ jet data, PDFs can be
derived which describe both data sets. An unambiguous search for
deviations, however, requires an independent measure of the PDFs.
Prospects for improvements to the PDFs will be covered in later
sections. First we turn our attention to inclusive jet measurements
at a different beam energy.

\section{RATIO OF JET CROSS SECTIONS AT TWO BEAM ENERGIES}

\subsection{Inclusive Jet Cross Sections, Scaling, and the
Ratio of Dimensionless Cross Sections}

An alternative way to test QCD is to compare the inclusive jet
cross sections measured  at widely separated center--of--mass energies. The
hypothesis of ``scaling'' predicts that the dimensionless, scaled
jet cross section,
\begin{center}
$E_T^4 (Ed^{3}\sigma / dp^{3})$,
\end{center}
will be independent of \rs~as a function of the
dimensionless
variable $x_T = 2E_T/\sqrt{s}$.
This can be written in terms of the experimentally measured quantities as,
\begin{center}
$\sigma_d = ( E_T^3 / 2\pi) d^{2}\sigma / d\ET d\eta$.
\end{center}
QCD predictions depend on the energy scale, or $Q^2$, of the
interactions and thus suggest that the cross sections should not
``scale''. The running of the strong coupling and the
evolution of the PDFs are manifestations of this energy (or scale)
dependence of the predictions.

\begin{figure}
\centerline{
  \epsfxsize=3.5in
  \epsfysize=2.5in
  \epsffile{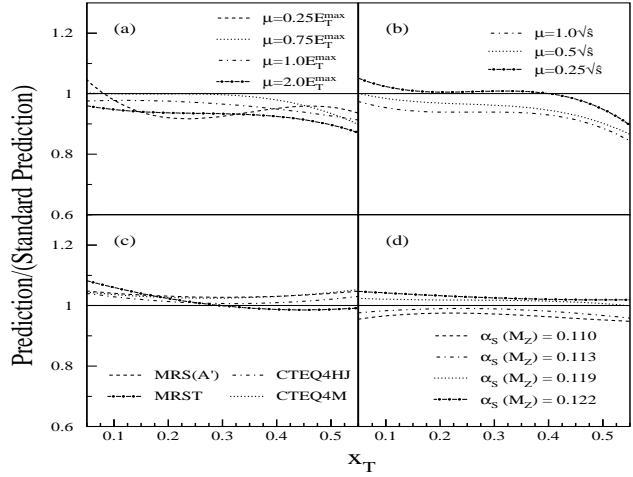}}
\caption{The difference between alternative predictions and the reference
prediction ($\mu = 0.5 E^{max}_{T}$, CTEQ3M) of the ratio of
inclusive jet cross sections at \rs = 630 and 1800 GeV for
$\eta_{jet} \leq 0.5$.  The alternative predictions are for the
choices (a) $\mu$ = 0.25, 0.75, 1.0 and 2.0 $\times E^{max}_{T}$,
(b) $\mu$ = 0.25, 0.5, and 1.0 $\times \rs$, (c) CTEQ4M, CTEQ4HJ,
MRS(A\' ), and MRST, and (d) calculations with the CTEQ4A series of
PDFs (which vary $\alpha_s$) compared with the calculation using
CTEQ4M.}
\label{TRATIOUNC}
\end{figure}

Measurement of the ratio of the scaled jet
cross sections from two different center--of--mass energies
but in the same experimental apparatus, provides a test of QCD in which
many theoretical and experimental uncertainties cancel.
Figure~\ref{TRATIOUNC} shows the ratio of the scaled cross
sections, $\sigma_d^{630}/\sigma_d^{1800}$, calculated with JETRAD.
The top two panels in the figure show a 10\% variation in the ratio
below $x_{T}$ = 0.4 (jet \ET = 360 GeV at \rs=1800 GeV) due to the
choice of scale.  The bottom two panels of Fig.~\ref{TRATIOUNC}
demonstrate that PDF choices produce variations below 10\%.

\subsection{Jet Production at \Ge{630}}

In Dec.~1995 CDF and D0 each collected $\approx$600 nb$^{-1}$ at
$\sqrt{s}$=630 GeV (beam energies of 315 GeV). A primary motivation
for this data run was the hope that the inclusive jet cross section
at this reduced energy and the ratio relative to 1800 GeV would
shed light on the deviations seen with the Run 1A 1800 GeV cross
section.  The analysis of the 630 GeV data follows an identical
path to the analysis described above for the 1800 GeV data. The
preliminary measurements from CDF~\cite{cdf630} and D0~\cite{D0630}
of the inclusive cross section at $\sqrt{s}=$630 GeV are shown in
Fig.~\ref{D0CDF630}. The CDF measurement is for the region $0.1
\leq \AEA \leq 0.7$ while the D0 result is for the region $\AEA
\leq 0.5$. The figure shows the percent difference between the data
and the associated theory prediction. The NLO theoretical
predictions used the MRSA' ~PDF and $\mu = 0.5 E_{T}^{jet}$. The
two measurements are in agreement above 80 GeV, but some
discrepancy exists near and below 60 GeV. The discrepancies are
within the 20-30\% systematic uncertainties reported by D\O\ and
represented by the shaded boxes. With regards to theory, the QCD
prediction is larger, though not significantly, than the data for
\ET less than 80 GeV. Note that in this $E_T$ range, results at
$\sqrt{s}$=1800 GeV show good agreement between data and
theoretical predictions.

\subsection{The Scaled Cross Sections}

\begin{figure}
\centerline{
  \epsfxsize=3.0in
  \epsfysize=2.5in
  \epsffile{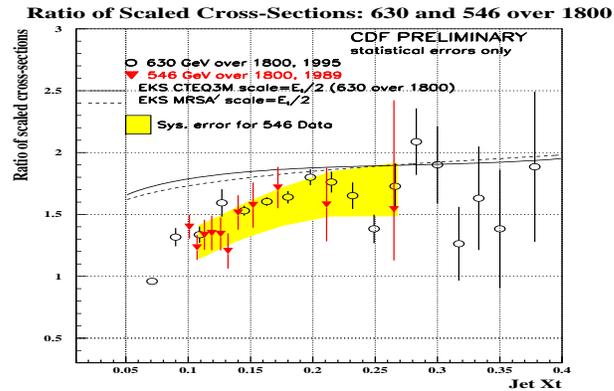}}
\caption{CDF scaling result from Tevatron runs at \rs = 546 and 1800 GeV
in 1989,
compared to the preliminary result from the Run 1B runs at 630 GeV
and 1800 GeV.}
\label{XT546_630}
\end{figure}

Figure~\ref{XT546_630} shows the preliminary ratio
of the scaled cross sections~\cite{cdf630} from CDF, $\sigma_{d}^{630}/
\sigma_{d}^{1800}$ along with the previous CDF
result~\cite{CDF-XT-ANALYSIS} for
$\sigma_{d}^{546}/\sigma_{d}^{1800}$ from much smaller data
samples.
The shaded band indicates the systematic uncertainty on the 546/1800 ratio;
the systematics on the 630/1800 GeV ratio are similar, but not yet final.
Good agreement between the results is observed.
The 546 GeV result ruled out ``scaling'' at the 95\% confidence
level and a disagreement with QCD predictions was observed at low
$E_T$.

\begin{figure}
\centerline{
\epsfysize=3.5in
\epsfxsize=3.5in
\leavevmode\epsffile{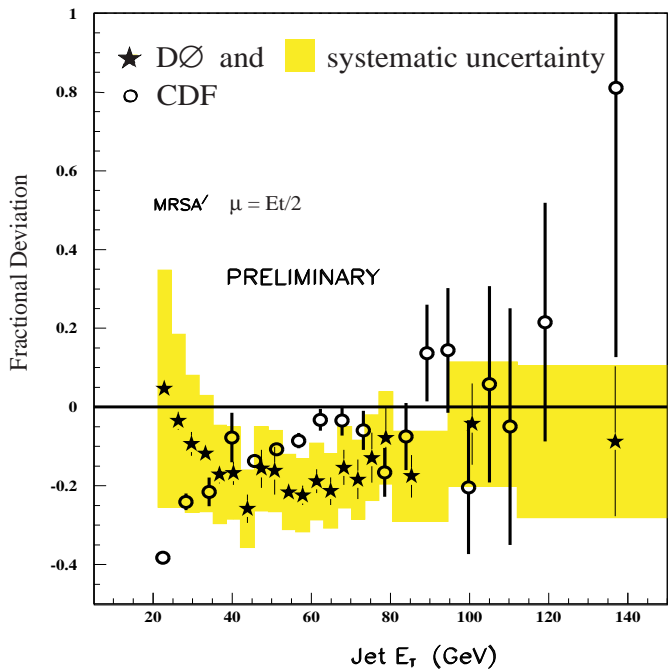}}
\caption[]{Preliminary D\O\ and CDF \rssps ~cross sections
compared to NLO QCD predictions. The shaded boxes represent
the D\O\ systematic errors.}
\label{D0CDF630}
\end{figure}

\begin{figure}
\centerline{
\epsfysize=3.5in
\epsfxsize=3.5in
\leavevmode\epsffile{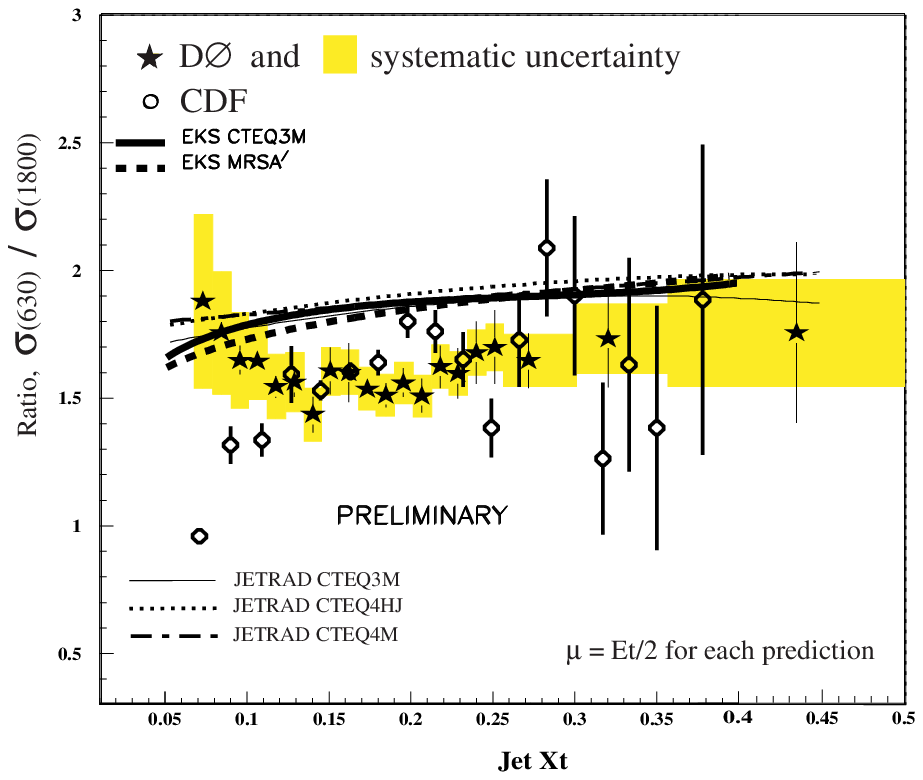}}
\caption[]{Preliminary D\O\ and CDF cross section
ratios for \rssps~to \rstev~compared to NLO QCD predictions. The
shaded boxes represent the D\O\ systematic errors.}
\label{D0CDFXT}
\end{figure}

Fig.~\ref{D0CDFXT} shows the preliminary cross section ratios from
D0~\cite{D0630} and CDF compared to
a variety of theoretical predictions.
Many of the energy
scale and luminosity uncertainties cancel in the ratio.
The uncertainty on the D0
ratio, shown by the shaded boxes, is about 7\%, much less than the
15-30\% uncertainty on the cross sections, and about a factor of
two better than previous ratio measurements.  The CDF and
D0 ratios are consistent with each other for $x_{T}\geq0.1$, but
some difference may exist for $x_{T}\leq0.1$. The discrepancy
between the two measured ratios is a reflection of the discrepancy
in Fig.~\ref{D0CDF630}. The significance of the difference must
await completion of the CDF systematic uncertainties. The theory is
roughly 20\% higher than the ratio measured by D0.
Figure~\ref{D0CDFXT} also shows three NLO QCD \JETRAD predictions
for the ratio using $\mu=0.5 E_{T}^{\rm max}$,
$\cal{R}_{\rm{sep}}$=$1.3$ and different PDFs. In addition,
two NLO predictions from the \EKS program using CTEQ3M and MRSA\'
~and $\mu=0.5 E_{T}$ are shown. Notice that the variation in
the predicted ratios is very small.

The preliminary
inclusive jet
cross sections at $\sqrt{s}=$630 GeV
are not well described by NLO QCD calculations.  Quantitative results on
these comparisons
await determination of final experimental uncertainties.
The ratio of inclusive cross sections is also in mild
disagreement with the theory.
With a larger data sample these measurements could place
constraints on the high $x$ behavior of the PDFs while using relatively
low $E_T$ jets.
An additional run at a similar beam energy should be considered for Run 2.
We now
turn to a discussion of a different technique of probing high $x$ behavior:
the study of the correlations between the leading two
jets resulting from a $p\bar{p}$ collision.

\section{DIJET DIFFERENTIAL CROSS SECTIONS AT LARGE RAPIDITY}

\subsection{Introduction}

Measurements of the dijet differential cross sections in different
rapidity regions can provide additional information and constraints
on the QCD predictions. By restricting the $E_T$ and rapidity of
the leading two jets in the events, different regions of $x$ and
$Q^2$ can be probed.  This may permit a more direct measure of the
proton PDFs at Tevatron energies.  For instance at LO,
an \ET = 90 GeV, $\eta = 0$ ($\theta$=90$^o$) jet must be balanced
by a second
\ET = 90 GeV jet.  If the second jet is at $\eta=0$, both jet
energies equal 90 GeV and so $x_1 = x_2 = 0.10$.  However if the
second jet is more forward,
since $E_T = E\sin\theta$ and $\sin\theta$ is smaller,
the jet energy and its fraction of the
initial hadronic momentum must increase to maintain \ET = 90 GeV.
At LO the parton momentum fraction $x$ is related to the $E_T$ and
$\eta$ of the two jets by the equations:
\[
x_1 = \frac{E_T}{\sqrt{s}} ( {\rm e}^{\eta_1} + {\rm e}^{\eta_2} ),
\;\;\;\;x_2 = \frac{E_T}{\sqrt{s}}
( {\rm e}^{-\eta_1} + {\rm e}^{-\eta_2} ).
\]
For \ET = 90 GeV jets at $\eta = 0$ and $2$, $x = 0.06$ and
$0.42$. For multi--jet production the calculations of $x$ are
generalized to
\[
x_1 = \frac{1}{\sqrt{s}} \sum_{i} E_{Ti} {\rm e}^{\eta_{i}},
\;\;\;\;x_2 = \frac{1}{\sqrt{s}}\sum_{i} E_{Ti} {\rm e}^{-\eta_{i}} .
\]
where ``$i$'' runs over all jets in an event.

Experimentally, the differential dijet cross sections are most
conveniently given by
\[\frac{d^{3}\sigma}{ d\ET d\eta_{1} d\eta_{2}} ,\]
where $\eta_1$ and $\eta_2$ are the pseudorapidities of the leading
two jets.  As with the inclusive cross section, the dijet
differential cross sections are integrated over a range of
pseudorapidity.  The cross sections are also integrated over a
range of $E_T$, but here there is even more freedom in the
definition of $E_{T}$: the leading jet \ET, the average of the
leading two jet $E_{T}$'s, or both leading jet $E_{T}$'s (two
entries into the cross section, one for the \ET of each of the
leading two jets). The specific choice depends upon the
experimental conditions. Although they are still preliminary we
present the dijet differential cross sections here in order to
demonstrate the potential and strength of these measurements. Once
the systematic uncertainties are well understood these measurements
will strongly constrain the PDFs.

\subsection{The CDF Measurement}

\begin{figure}
\centerline{
  \epsfxsize=4.0in
  \epsfysize=4.5in
  \epsffile{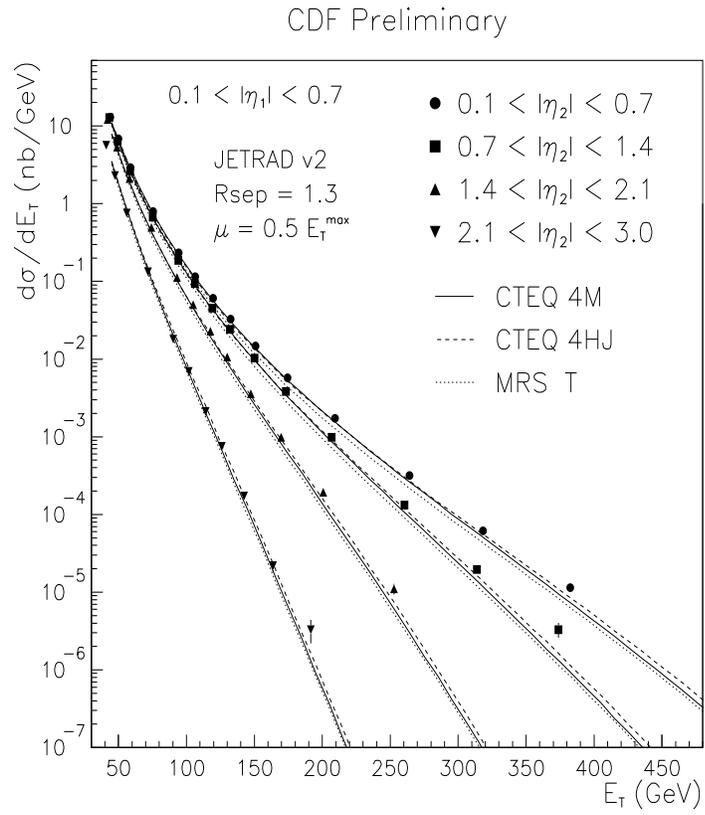}}
\caption{CDF cross sections for central jets with the second jet
in different rapidity intervals.}
\label{CDFDIFLOG}
\end{figure}

\begin{figure}
\centerline{
  \epsfxsize=4.0in
  \epsfysize=4.5in
  \epsffile{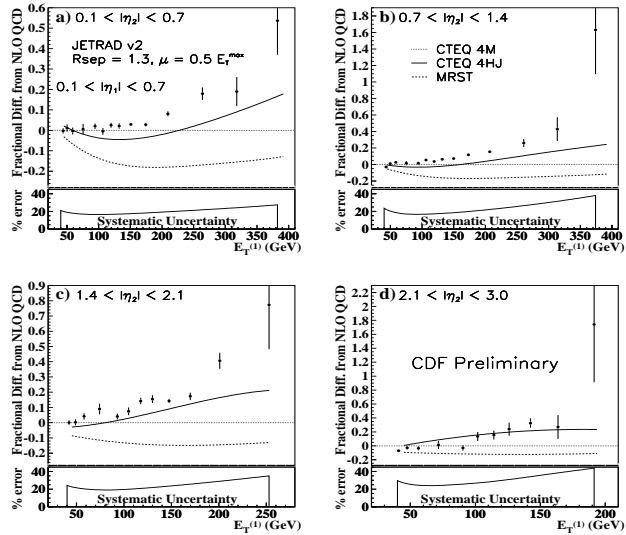}}
\vspace{-1.3in}
\caption{Percent difference between preliminary CDF data (points)
and a NLO QCD prediction using the \JETRAD program, CTEQ4M, $R_{sep}$=1.3
and $\mu$=$E_T^{max}/2$.  The percent difference between predictions using CTEQ4M
and predictions using MRST and  CTEQ4HJ are also shown.
The four plots represent the
cross section as a function of the central jet $E_T$ when the
rapidity of the second jet is restricted to different ranges. The
curve in the lower box represents the quadrature sum of the
correlated systematic uncertainties.}
\label{cdfdiflin}
\end{figure}

In the CDF Run 1B measurement, the two highest $E_T$ jets are
identified and one is required to be in the central
(0.1$\leq|\eta|\leq$0.7) region. Because the central region has the
smallest energy scale uncertainty, the central jet is used to
measure the $E_T$ of the event. The other jet, called the ``probe"
jet, is required to have $E_T>$10 GeV and to fall in one of the
$\eta$ bins: 0.1$\leq|\eta|\leq$0.7, 0.7$\leq|\eta|\leq$1.4,
1.4$\leq|\eta|\leq$2.1, 2.1$\leq|\eta|\leq$3.0. There are no
restrictions on the presence of additional jets.
Figure~\ref{CDFDIFLOG} shows the preliminary cross
section~\cite{CDFNEW} in the individual $\eta$ bins as a function
of the central jet $E_T$. \JETRAD is used for the theoretical
predictions with scale $\mu$ = 0.5 $E_{T}^{max}$ and $R_{sep}$=1.3.
The data
are compared to the predictions using three parton distribution
functions, CTEQ4HJ, MRST and CTEQ4M. The statistical uncertainty is
shown on the points. Figure~\ref{cdfdiflin} shows the percent
difference between the data and theory (CTEQ4M) as a function
of the central jet $E_T$ for each rapidity bin.
The additional curves represent the percent difference from the prediction
using CTEQ4M for predictions using MRST and CTEQ4HJ.
The systematic uncertainty on the measurement is
evaluated in a manner similar to the inclusive jet cross section.
The quadrature sum of the systematic uncertainties is shown in the
box below the points. The high rapidity bins reach $x$ of
roughly 0.6--0.7, while the $E_T$ of the jets is in the 100~--~200
GeV range. The excess over predictions using CTEQ4M
observed in the inclusive
cross section at high $E_T$ is seen in all four rapidity bins. This
suggests that it is not a function of the jet $E_T$, but rather a
function of $x$ and is thus related to inadequacies in the PDFs.
The improved agreement in all bins when CTEQ4HJ is used
is similarly suggestive.

\subsection{The D0 Measurement}

The D0 Run 1B (92 \ipb) measurement organizes the dijet
differential cross section according to the rapidity of both
leading jets.  The rapidities are divided into four same--side (SS)
bins where $\eta_1 \sim \eta_2$ and four opposite--side (OS) bins
where $\eta_1 \sim -\eta_2$. The bins and approximated $x$ ranges sampled
(assuming a L0 process and the observed \ET ranges in each bin) are
listed in Table~\ref{TAB:D0XSAMPLE}.  The eight cross sections
are all plotted versus jet $E_T$.  Here each event has two
entries, one for the \ET of each of the leading two jets.  As indicated in
the table, the low rapidity measurements can provide confirmation
of previous PDF measurements extrapolated from low $Q^2$.  On the
other hand, the large rapidity, SS events probe much larger $x$
values.

\begin{table}
\begin{center}
\caption{ $x$ ranges of SS and OS dijet differential cross sections. }
\label{TAB:D0XSAMPLE}
\vspace{0.2cm}
\begin{tabular}{|c|c|c|c|}
\hline
$\eta$ bin      & Topology   & $x_{min}$ & $x_{max}$ \\ \hline
0.0-0.5         &    SS      &    0.04   &    0.53   \\
0.0-0.5         &    OS      &    0.07   &    0.42   \\ \hline
0.5-1.0         &    SS      &    0.03   &    0.73   \\
0.5-1.0         &    OS      &    0.08   &    0.45   \\ \hline
1.0-1.5         &    SS      &    0.02   &    0.81   \\
1.0-1.5         &    OS      &    0.10   &    0.44   \\ \hline
1.5-2.0         &    SS      &    0.01   &    0.80   \\
1.5-2.0         &    OS      &    0.16   &    0.46   \\ \hline
\end{tabular}
\end{center}
\end{table}

\begin{figure}
\centerline{
  \epsfxsize=4.0in
  \epsfysize=2.5in
  \epsffile{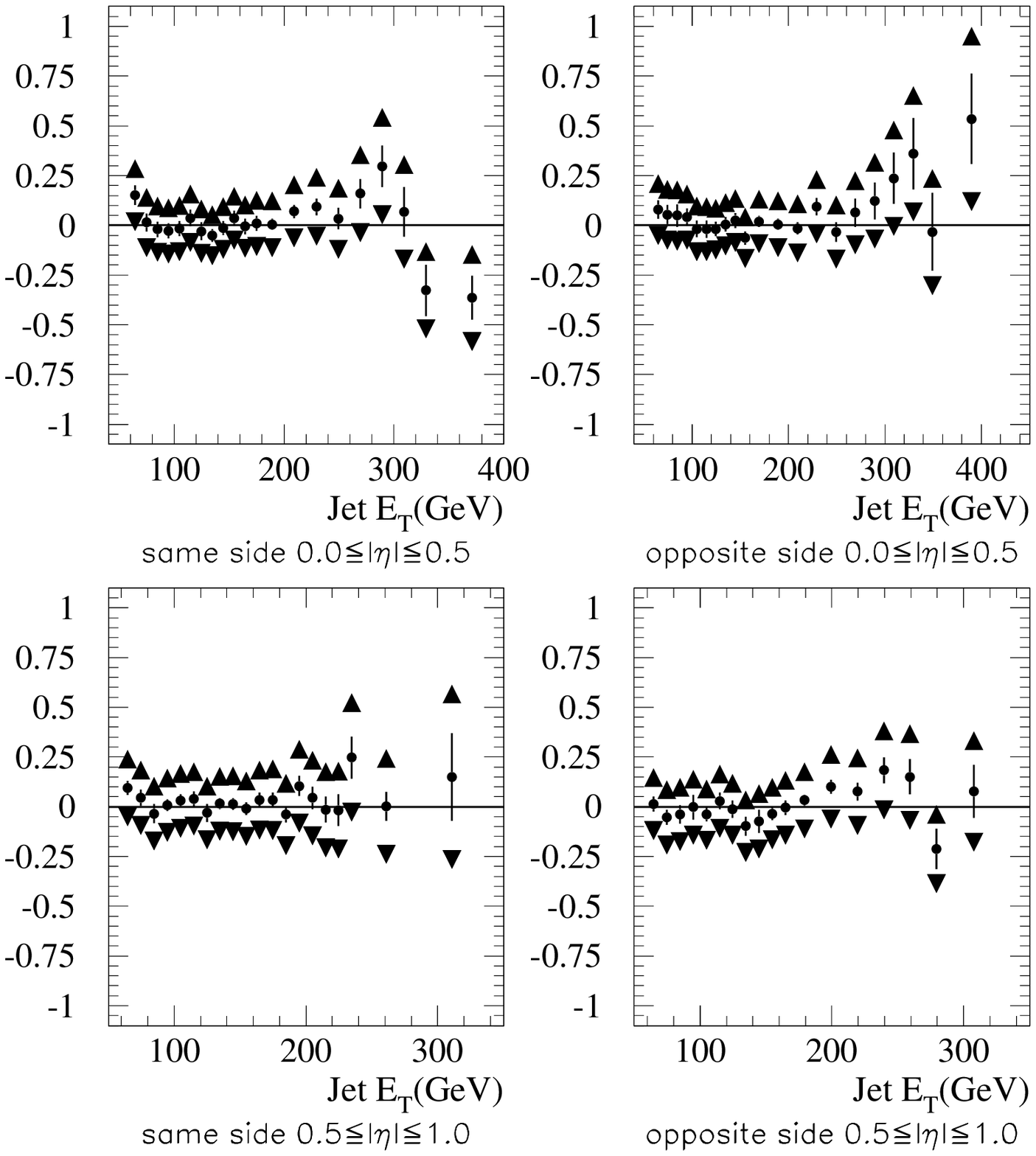}}
\caption{D0 same--side and opposite--side cross sections compared to
\JETRAD with $R_{sep}$=1.3, $\mu=E_T^{max}/2$ and CTEQ3M. The
two leading jets are restricted to different $\eta$ regions.  See the
text for details.}
\label{D0TRIPLE1}
\end{figure}

\begin{figure}
\centerline{
  \epsfxsize=4.0in
  \epsfysize=2.5in
  \epsffile{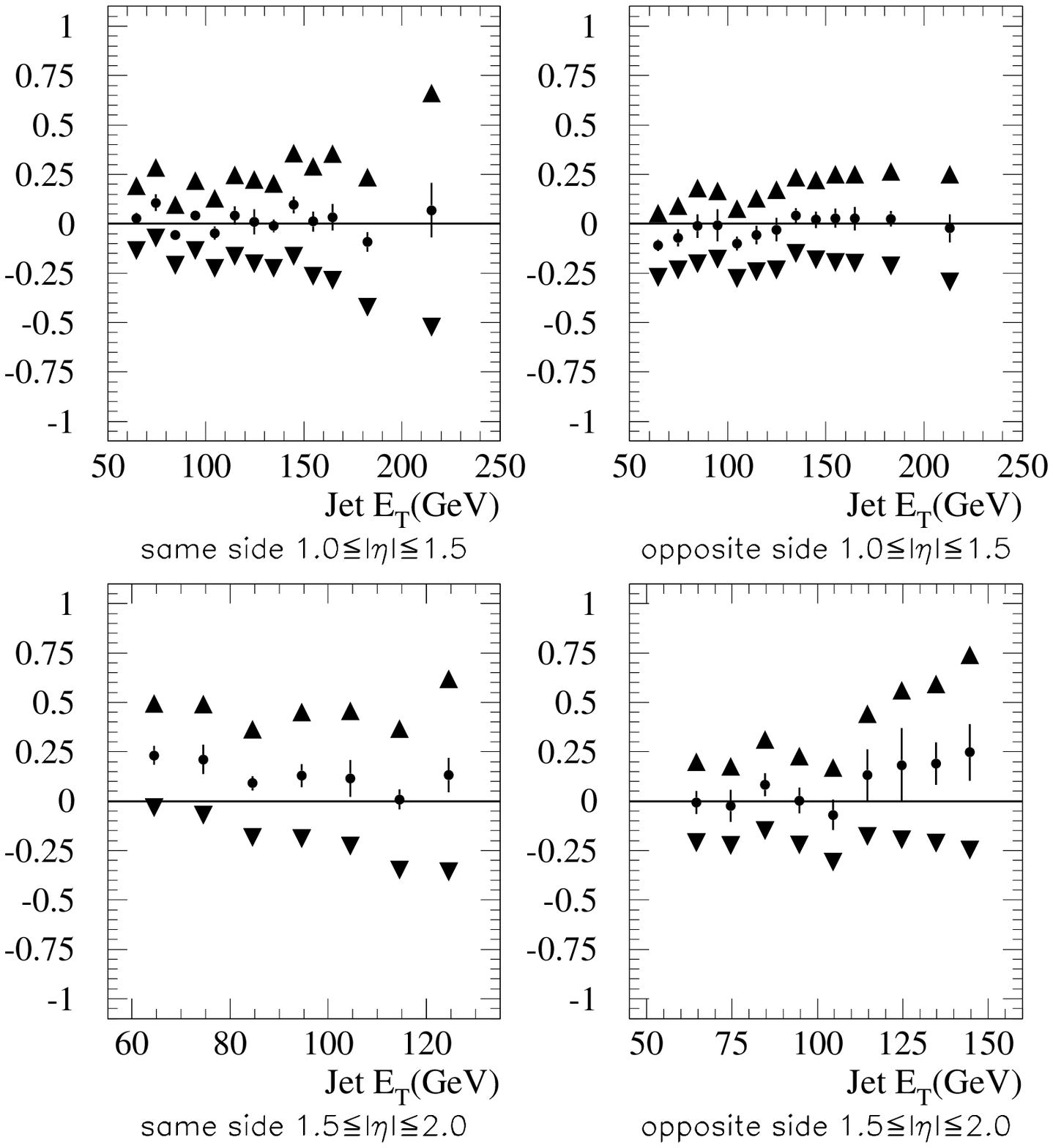}}
\caption{D0 same--side and opposite--side cross sections compared to
\JETRAD with $R_{sep}$=1.3, $\mu=E_T^{max}/2$ and CTEQ3M. The
two leading jets are restricted to different $\eta$ regions. See the
text for details.}
\label{D0TRIPLE2}
\end{figure}

The D0 analysis follows the inclusive analysis very closely. Jet
and event selection, energy correction, and resolution unsmearing
are all quite similar.  An additional correction for vertex
resolution is also important in the very forward--backward rapidity
bins. The cross section uncertainties are quite similar to the
inclusive cross section for the four rapidity bins limited by $\eta
= 1.0$. In the high $\eta$ bins the systematic errors are approximately
doubled.  Figs.~\ref{D0TRIPLE1} and \ref{D0TRIPLE2} show the
fractional difference between data and theory for all eight
bins~\cite{CDFNEW, D0TRIPLE}. The theoretical prediction is from
\JETRAD with CTEQ3M and $\mu = 0.5 E_T^{max}$. The
bars on the data represent the statistical uncertainty and the
outer symbols the total uncertainty.  Note there is good agreement
over all rapidity for $x=0.01$ to $0.80$.  Significantly, this PDF
includes no collider jet data.  Although not shown, the agreement
with CTEQ4M is also reasonable.

\subsection{Prospects}

The differential cross sections show great promise for constraining
the PDFs.  In all rapidity bins the
CDF data appears to prefer CTEQ4HJ over CTEQ4M or MRST,
while the D0 data seems to be good agreement with the predictions using CTEQ3M.
This apparent disagreement mirrors the situation in the
inclusive cross sections.  However, a firm statement on the
agreement or disagreement of the two data sets is
obscured by the different techniques and the
current lack of quantitative comparisons between data and theory.

\section{DIJET MASS AND ANGULAR DISTRIBUTIONS AT 1800 TEV}

\subsection{Introduction}

The LO cross section for $p\bar{p} \rightarrow$
$jet_1$ + $jet_2$ + $X$ events (where $jet_1$ and $jet_2$ are the
leading two jets) can be completely described in terms of three
orthogonal center--of--mass variables. These are cos $\theta^*$,
where $\theta^*$ is the center--of--mass scattering angle between
the two leading jets, the boost of the dijet system
\mbox{$\eta_{boost} = (\eta_1 + \eta_2)/2$,}
and the dijet mass, $M_{jj}$ as
follows~\cite{ELLISNPB}:
\[\frac{d^3\sigma}{d\eta_{boost}dM_{jj}dcos\theta^*} =
\frac{\pi \alpha_s^2(Q^2)}{2s^2}
(2M_{jj})
\sum_{1,2} \frac{f(x_1,Q^2)}{x_1}\frac{f(x_2,Q^2)}{x_2}
|m_{12}|^2,
\]
where $\alpha_s^2(Q^2)$ is the strong coupling strength,
$|m_{12}|^2$ is the hard
scattering matrix element, $x_1$ ($x_2$) is the fraction of the
proton (antiproton) momentum carried by the parton and $f(x_1,Q^2)$
is the parton momentum distribution.
Typically the dijet mass is derived from measured variables
such as \ET, $\eta_{jet}$, and $\phi_{jet}$.  In the case of higher
order processes, where more than two jets are produced, the mass is
calculated using the two highest \ET jets in the event and
additional jets are ignored.

Integration of the general dijet cross section over boost and
production angle leads to the dijet mass spectrum.  The spectrum is
a useful test of QCD sensitive to the PDFs.  On the other hand
integration over mass and boost leads to the dijet angular
distribution -- a marvelous test of the hard scattering matrix
elements almost totally insensitive to the PDFs. Comparisons of the
angular distributions and certain mass spectra ratios to
theoretical predictions can establish stringent limits on the
presence of conjectured quark constituents. We turn now to a discussion
of the these spectra and compositeness limits.

\subsection{The Mass Distributions}

\begin{figure}
\centerline{
  \epsfxsize=4.0in
  \epsfysize=2.5in
  \epsffile{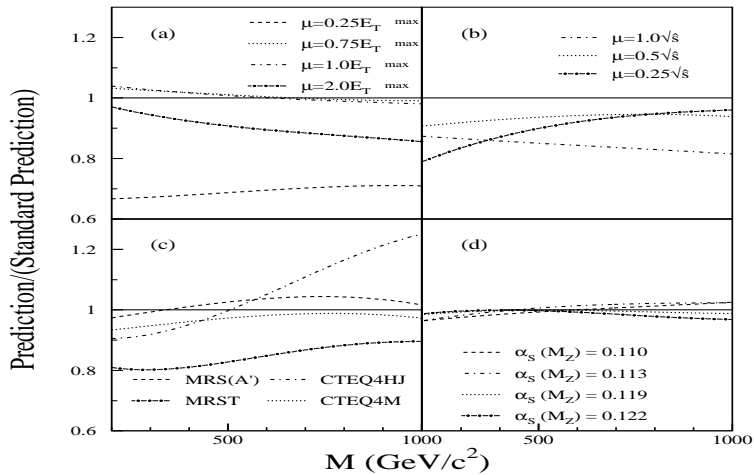}}
\caption{The difference between alternative predictions and the reference
prediction ($\mu = 0.5 E^{max}_{T}$, CTEQ3M) for the
inclusive dijet mass cross section at \rs = 1800 GeV for
$\eta_{jet} \leq 1.0$.  The alternative predictions are for
(a) $\mu$ = 0.25, 0.75, 1.0 and 2.0 $\times E^{max}_{T}$,
(b) $\mu$ = 0.25, 0.5, and 1.0 $\times \rs$, (c) CTEQ4M, CTEQ4HJ,
MRS(A\' ), and MRST, and (d) calculations with the CTEQ4A series of
PDFs (which vary $\alpha_s$) compared with the calculation using
CTEQ4M.}
\label{mass_theory_uncertainties}
\end{figure}

At LO, where only two jets are produced, the dijet invariant
mass is given by \[ M_{jj}^2 = \hat{s} = x_1x_2s \] where $\hat{s}$ is
the center of mass energy of the interacting partons and $s$ the
total center--of--mass energy.  Since the dijet mass represents the
center--of--mass energy of the interaction it directly probes the
parton distribution functions. Experimentally, the dijet
mass cross section is given by
\[\frac{d^{3}\sigma}{ dM_{jj} d\eta_{1} d\eta_{2}} ,\]
where $\eta_1$
and $\eta_2$ are the pseudorapidities of the jets.  As with the
inclusive jet cross section the dijet mass cross section is
integrated over a range of pseudorapidity. For example,
Fig.~\ref{mass_theory_uncertainties} shows NLO QCD dijet
predictions for $|\eta_{jet}| \leq 1.0$ relative to a reference
prediction.  There are 20--30\% variations due to the
scale and to the PDFs.

\begin{figure}
\centerline{
  \epsfxsize=4.0in
  \epsfysize=2.5in
  \epsffile{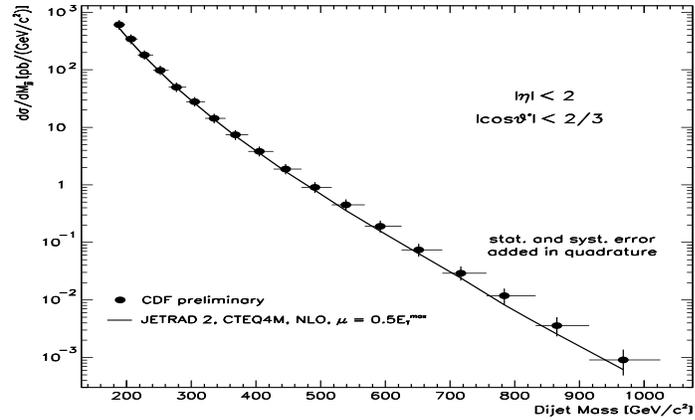}}
\caption{Dijet mass as measured by CDF compared to the NLO prediction
from \JETRAD with CTEQ4M and $\mu=E_T^{max}/2$ and $R_{sep}$=1.3.
The error bars
indicate the quadrature sum of the statistical and systematic
uncertainties}
\label{cdfdjmlog}
\end{figure}

\begin{figure}
\centerline{
  \epsfxsize=4.0in
  \epsfysize=2.5in
  \epsffile{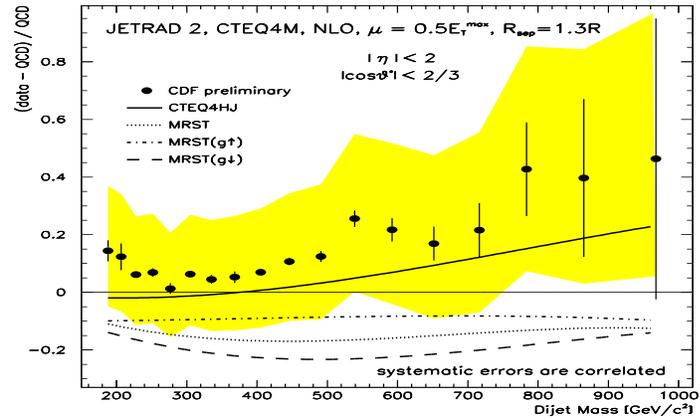}}
\caption{
CDF data compared to predictions from the \JETRAD program for
CTEQ4M and $\mu=E_T^{max}/2$ and $R_{sep}$=1.3 (full circles).
Predictions using
other PDFs and $\mu=E_T^{max}/2$ are also compared to CTEQ4M: MRST
(dotted), CTEQ4M with $\mu=E_T^{max}$ (dashed), and CTEQ4HJ (solid).
The error bars indicate the statistical errors. The shaded area
represents the combined systematic uncertainty.}
\label{cdfdjmlin}
\end{figure}

The D0 and CDF collaborations have both measured the dijet mass
spectrum with Run 1B data samples. The CDF measurement uses the
four--vector definition for the dijet mass
$M_{jj}=\sqrt{(E_1+E_2)^2-(P_1 +P_2)^2}$ where $E$ and $P$ are the
energy and momentum of a jet and allows the rapidity of the jets to
extend to $|\eta_{jet}|<2.0$. To ensure good acceptance, the CDF
analysis also imposes a cut on $|cos\theta^*|<2/3$, where
$\theta^*$ is the scattering angle in the center--of--mass frame.
Figure~\ref{cdfdjmlog} shows the CDF preliminary data compared to
\JETRAD predictions~\cite{CDFNEW}. The error bars represent the
quadrature sum of the statistical and systematic uncertainties. The
data and \JETRAD predictions using CTEQ4M, $\mu=0.5
E_T^{jet}$ and $R_{sep}=1.3$ are in good agreement.
Figure~\ref{cdfdjmlin} shows the percent difference between the
data and the theoretical prediction. The shaded band shows the
preliminary estimate of the systematic uncertainties. These
uncertainties are derived in a manner similar to the inclusive jet
cross section measurement with additional contributions for jets
outside the central region. The percent difference between the
default prediction and those using other PDFs are also shown and
are all consistent with the data given the systematic errors.
However, as with the inclusive and dijet cross section
measurements, CTEQ4HJ seems to provide the best agreement with the
data in the high $E_T$ region.

\begin{figure}
\centerline{
  \epsfxsize=4.0in
  \epsffile{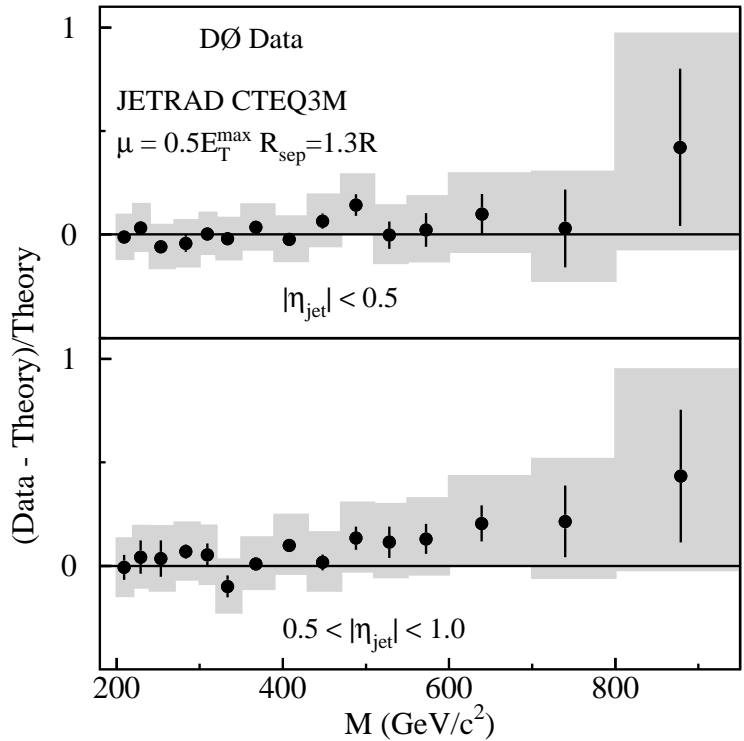}}
  \caption{The difference between the D0 dijet mass data and the
   prediction (\JETRAD) divided by the prediction for
   $|\eta_{jet}| \leq 0.5$ and $0.5 < |\eta_{jet}| < 1.0$. The
   solid circles represent the comparison to the
   calculation using CTEQ3M with $\mu = 0.5 \ET^{max}$.
   The shaded region represents the $\pm$$1\sigma$ systematic uncertainties.}
  \label{D0DIJETETA}
\end{figure}

\begin{figure}
\centerline{
  \epsfxsize=4.0in
  \epsffile{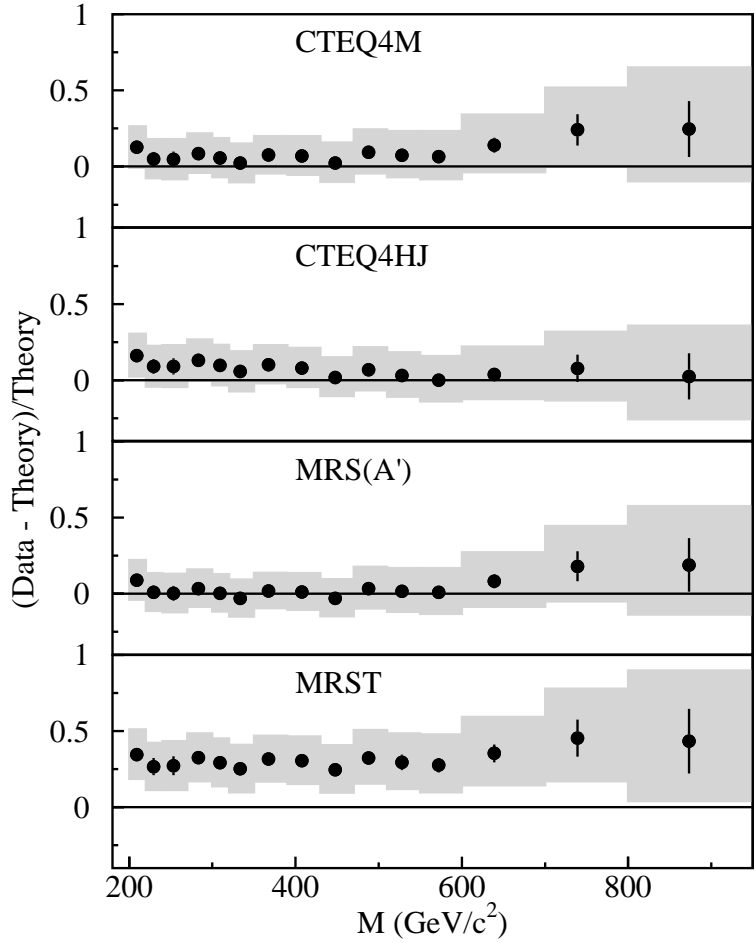}}
  \caption{For $|\eta_{jet}| < $1.0, the difference between 
   the D0 dijet mass data and JETRAD predictions 
   divided by the predictions. 
   The calculations used $\mu =
   0.5 \ET^{max}$, $R_{sep}$=1.3 and CTEQ4M, CTEQ4HJ, MRS(A$^\prime$), and
   MRST. The error bars represent the statistical uncertainty on the data.
   The shaded region represents the $\pm1\sigma$ systematic
   uncertainties}
  \label{D0DIJETPDF}
\end{figure}

\begin{figure}
\centerline{
  \epsfxsize=4.0in
  \epsfysize=2.5in
  \epsffile{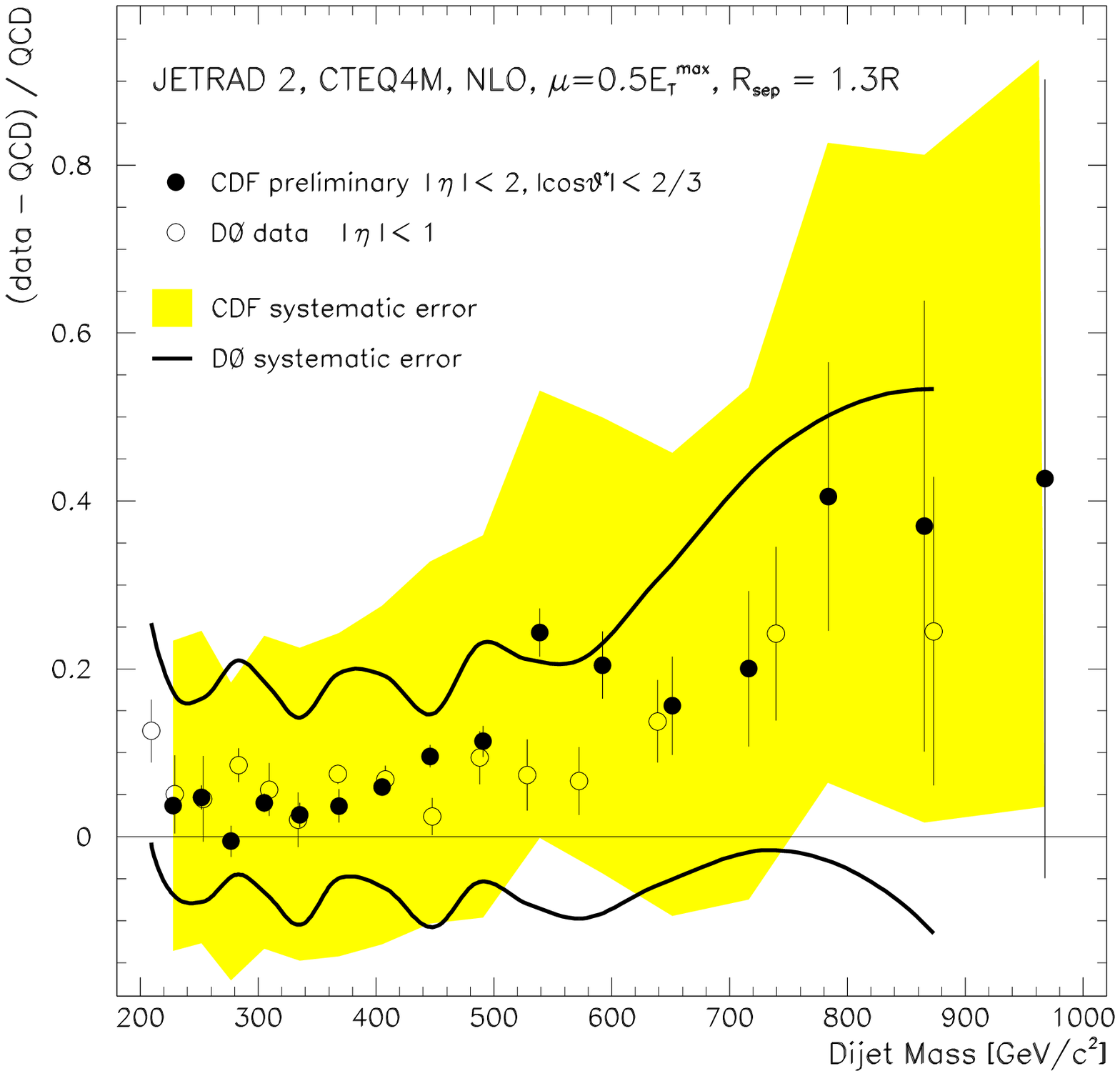}}
\caption{
D0 (open circles) and preliminary CDF (full circles) dijet mass
results compared to predictions from the \JETRAD program for CTEQ4M
and $\mu=E_T/2$.}
\label{djmcdfd0}
\end{figure}

D0 has performed the dijet mass measurement in two rapidity
regions: central, $|\eta_{jet}|<0.5$, and more forward,
$|\eta_{jet}|<1.0$, $|\Delta\eta_{jet}|<1.6$~\cite{D0DIJETMASS}.
The dijet mass at D0 is defined assuming massless jets:
$M_{jj}=\sqrt{2E_T^1E_T^2(cosh(\Delta \eta) - cos(\Delta \phi))}$,
where  $\Delta \eta$ and $\Delta \phi$ are the rapidity and
azimuthal separation of the two jets. Figure~\ref{D0DIJETETA} shows
the difference between the data and theory divided by the theory.
As with the D0 inclusive jet measurement there is good agreement
with theory, well within systematic uncertainties. At larger
rapidity the data shows a tendency to be slightly above the
predictions at high $E_T$. Data--theory $\chi^2$ comparisons
similar to those described in Section 4 yield large probabilities.
As Fig.~\ref{D0DIJETPDF} illustrates there is sensitivity to the
PDF choice but unfortunately its significance is obscured by the
systematic uncertainties. As with the inclusive jet cross section
these uncertainties are highly correlated. In Fig.~\ref{djmcdfd0},
the D0 $\AEA \leq 1.0$ result is compared to the CDF result and to
predictions with CTEQ4M.  The kinematic cuts of the two analyses
have a significant overlap; 59\% of the CDF sample has the two
leading jets within $\AEA \leq 1.0$. There is remarkable agreement
between these data samples over the full $E_T$ range.

\subsection{The Dijet Angular Distributions}

The dijet angular distribution measured in the center--of--mass is
sensitive primarily to the hard scattering matrix elements. In
fact, the distribution is unique among high $E_T$ measurements in
that it is almost independent of the PDFs. The shape of the angular
distribution is dominated by t--channel exchange and is nearly
identical for all dominant scattering subprocesses (e.g. $gg
\rightarrow gg$, $qg \rightarrow qg$, and $qq \rightarrow qq$). The
dijet angular distribution predicted by LO QCD (two jets only) is
proportional to the Rutherford cross section
\[\frac{d\sigma}{dcos\theta^*} \propto
\frac{1}{sin^4(\theta^*/2)} .\]
To flatten out the
distribution and facilitate the comparison to theory,
the variable transformation
$\chi = \frac{1 + |cos\theta^*|} {1 - |cos\theta^*|}$ is used
giving
\[\frac{d^{3}\sigma}{ dM_{jj} d\chi d\eta_{boost}} .\]

\begin{figure}
\centerline{
   \epsfxsize=4.0in
  \epsfysize=2.5in
  \epsffile{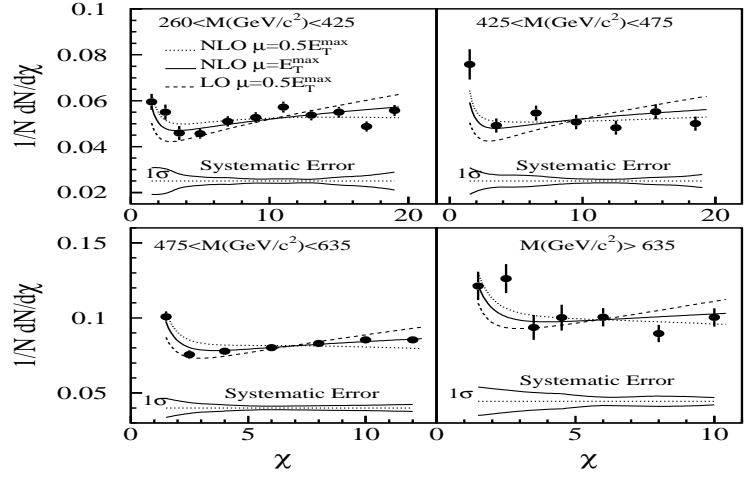}}
\caption{Dijet angular distribution as measured by D0 for
different mass ranges compared to LO and NLO QCD predictions.}
\label{d0dijang1}
\end{figure}

\begin{figure}
\centerline{
  \epsfxsize=4.0in
  \epsfysize=2.5in
  \epsffile{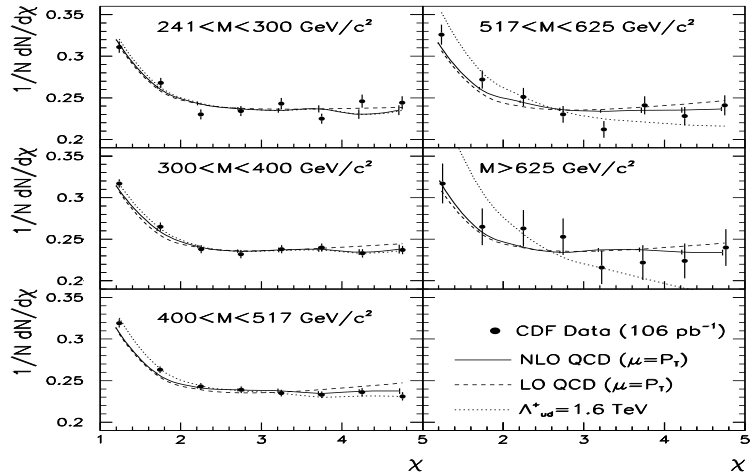}}
\caption{
Dijet angular distribution as measured by CDF for
different mass ranges compared to LO and NLO QCD predictions.}
\label{cdfdijang}
\end{figure}

For measurement of the dijet angular spectra this quantity is
integrated over pseudorapidity and mass regions and normalized to
the total number of events N in those regions. The normalization
reduces both experimental and theoretical uncertainties. Figure
\ref{d0dijang1} shows the $\frac{1}{N}\frac{ dN}{ d\chi}$
distributions from the D0 collaboration for four mass
ranges~\cite{D0ANGULAR98}. In all cases the jets are limited to
regions of full acceptance. The NLO predictions from \JETRAD with
$\mu= E_T^{max}$ provide the best agreement with the
shape of the data. The large $\chi$ acceptance of the measurement
allows discrimination between LO and NLO predictions.
Figure~\ref{cdfdijang} shows CDF data compared to QCD predictions
(\JETRAD) for different mass bins~\cite{CDFANGULAR96}. In this case
both LO and NLO QCD are in good agreement with the distribution.

\subsection{Compositeness and New Physics Limits}

The dijet mass and angular distributions are sensitive to new
physics such as quark compositeness.  In QCD parton--parton
scattering, the dominant exchange involves the t--channel.
This produces distributions peaked at small center--of--mass
scattering angles (near the beam axis in the lab) e.g. large $\eta$ and
$\chi$. In contrast, the compositeness model~\cite{COMP}
predicts a more isotropic angular distribution. Thus, relative to
QCD predictions, the contributions of composite quarks would
be most noticeable in the central region, near $\eta=0$ and $\chi$ = 1.

Compositeness signals may be parameterized by a mass scale
$\Lambda$ which characterizes the quark--substructure coupling.
Limits on $\Lambda$
are set assuming that $\Lambda \gg \sqrt{\rm \hat{s}}$, such that
the dominant force is still QCD.
The substructure coupling is
approximated by a four--Fermi contact interaction giving rise to an
effective Lagrangian~\cite{COMP}.  The Lagrangian contains
eight terms describing the coupling of left and right handed quarks
and antiquarks. Currently only the term describing the left handed
coupling of quarks and anti--quarks has been calculated, and this
term has an unknown phase.  Limits are reported for the case where
specific quarks or all quarks are composite with either
constructive interference ($\Lambda^{-}$) or destructive
interference ($\Lambda^{+}$). To set compositeness limits CDF and
D0 both use the NLO \JETRAD prediction times a ``k--factor'' from LO
QCD + compositeness ~\cite{CDFANGULAR96}.  NLO calculations with
compositeness are not available. Figure~\ref{cdfdijang} includes a
curve with a compositeness signal added. The additional contribution
at low $\chi$
is most pronounced in the highest dijet mass bins.

\begin{figure}
\centerline{
  \epsfxsize=3.5in
  \epsfysize=2.5in
  \epsffile{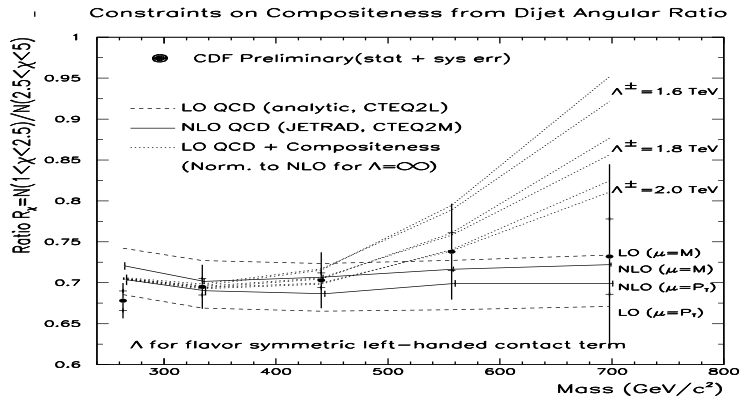}}
\caption{Dijet angular distribution as measured by CDF compared
to QCD and to QCD plus a term for composite quarks. Limits on
compositeness are derived from the dijet angular distribution ratio
of the number events below $\chi$=2.5 to the number between 2.5 and
5.}
\label{cdfang}
\end{figure}

\begin{figure}
\centerline{
  \epsfxsize=3.5in
  \epsfysize=2.5in
  \epsffile{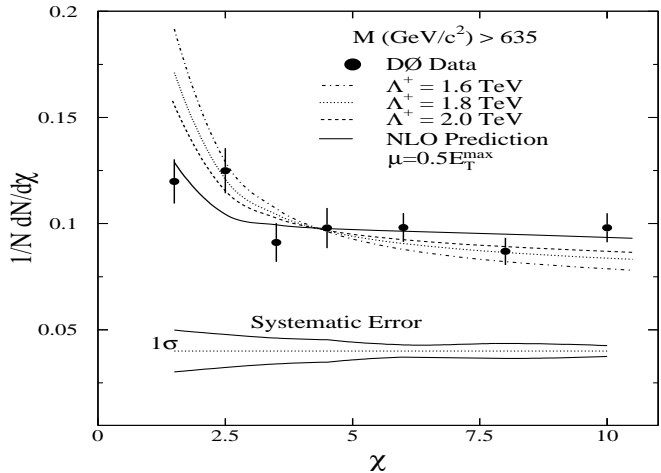}}
\caption{Dijet angular distribution as measured by D0
compared to predictions with additional contributions from
composite quarks. }
\label{d0dijang2}
\end{figure}

Both CDF and D0 set compositeness limits based on the ratio of
number of events at low $\chi$ to those in the high $\chi$ region.
Figure~\ref{cdfang} shows the CDF result for the ratio of the
number of events below $\chi$=2.5 and between $\chi$= 2.5 and 5.0
as a function of the dijet mass, along with curves which correspond
to the different values of $\Lambda$. Using this ratio, the CDF
data excludes at the 95\% C.L. a contact interaction scale of
$\Lambda^+_{ud} \le 1.6$~TeV and $\Lambda^-_{ud} \le 1.4$~TeV. For
a model where all quarks are composite $\Lambda^+ \le 1.8$~TeV and
$\Lambda^- \le 1.6$~TeV. Figure~\ref{d0dijang2} shows the angular
distribution as measured by D0 for very high dijet masses (greater
than 635 GeV/c$^{2}$) compared to predictions which include
composite quarks. Compositeness limits in this analysis are derived
from the ratio of the number of events above and below $\chi$=4.
The data excludes (at the 95\% CL) contact interaction scales with
$\mu=1.0E_{T}^{max}$: $\Lambda^-_{ud} \le 2.0$~TeV,
$\Lambda^- \le 2.2$~TeV, $\Lambda^+
\le 2.1$~TeV, and with $\mu=0.5E_{T}^{max}$:
$\Lambda^-_{ud} \le 2.2$~TeV, $\Lambda^- \le 2.4$~TeV, $\Lambda^+
\le 2.3$~TeV.

\begin{figure}
\centerline{
  \epsfxsize=4.0in
  \epsfysize=2.5in
  \epsffile{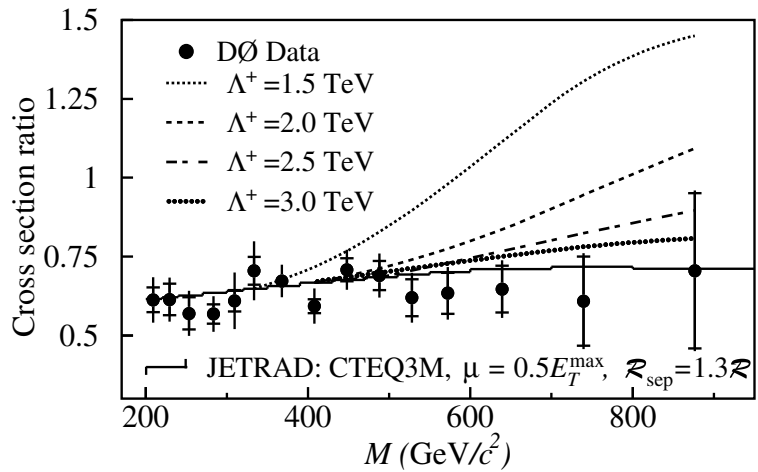}}
  \caption{The ratio of D0
   dijet mass cross sections for $|\eta_{jet}|  < 0.5$ and
   $0.5< |\eta_{jet}| <1.0$ for data (solid circles) and theory (various
   lines).  The error bars show the statistical and systematic
   uncertainties added in quadrature, and the crossbar shows the size of
   the statistical error.}
\label{D0MASSCOMP}
\end{figure}

The best limits on compositeness now come from recent dijet mass
measurements from the D0 collaboration~\cite{D0DIJETMASS} which
combine the sensitivity of the dijet mass distributions with the
PDF independence of the angular distributions.  In much the same
way jet production from a composite interaction increases the
$\chi$ spectrum at low values of $\chi$, the dijet mass spectrum
will increase at central rapidities relative to forward rapidities.
Thus the ratio of the central and forward dijet mass spectra will
be sensitive to $\Lambda$.  In addition, both theoretical and
experimental uncertainty are reduced in the ratio.
Fig.~\ref{D0MASSCOMP} shows the ratio of cross--sections for
$|\eta_{jet}| \leq 0.5$ and  $0.5 \leq |\eta_{jet}| \leq 1.0$ as a
function of dijet mass.  As indicated by the family of curves, the
compositeness model predicts changes in shape to this ratio at high
mass. The spectrum rules out quark compositeness at the 95\%
confidence level for $\Lambda^+$ below $2.7$~TeV and $\Lambda^-$
below $2.5$~TeV.

\section{CONCLUSIONS}

The inclusive jet and dijet cross sections provide fundamental
tests of QCD predictions at the highest jet $E_T$ and thus are the
deepest probes into the structure of the proton. With the increased
luminosity performance of the Tevatron, measurements of these cross
sections are no longer limited by statistical uncertainties. In
fact, the systematic uncertainties from the experimental
measurements and from the theoretical predictions are comparable in
size and are significantly larger than the statistical uncertainty
on all but the highest $E_T$ data points. Uncertainties in the PDFs
dominate the theoretical uncertainty, while uncertainties in the
jet energy scale dominate the experimental measurements.

The inclusive jet cross sections from the Tevatron have proved a
particularly interesting test of QCD.  The Run 1A inclusive jet
measurement showed disagreement with concurrent pQCD predictions at
the highest $x$ and $Q$. Two subsequent measurements, each using
five times the data, show mixed agreement between the data and
theory at high $E_T$.  One measurement is consistent with the Run
1A measurement, and can be described by QCD if the PDFs are
suitably modified.  The second Run 1B measurement is well described
by QCD with PDFs that either do or do not include high \ET jet
data. Further, within statistical and systematic errors, the three
measurements are compatible.  The preliminary measurements of the
inclusive jet cross section at $\sqrt{s}$=630 GeV and the ratio of
the cross sections show good agreement with previous results and
marginal agreement with QCD predictions.  Derivation of
quantitative results is in progress.  Unfortunately this sample is
too statistically limited to provide a constraint on the high $E_T$
behavior of the cross section at 1800 GeV.

The apparent excess of events at high $E_T$ observed in the
inclusive jet cross section from Run 1A created intense scrutiny of
the theoretical predictions.  As a result, there is now a better
understanding of the uncertainty in the theoretical predictions,
particularly for the parton distribution functions. 
The dijet mass
distributions and differential cross sections reflect a story
similar to the inclusive jet measurements. The results are
generally consistent with each other and with QCD, but with
hints that moderately forward regions indicate some high $x$
modifications of the PDFs might be appropriate. Incorporation of
this information into the global fitting procedures used to derive
the PDFs should be particularly helpful in reducing the uncertainty
associated with extrapolation from low energy DIS data to the
kinematic region covered by the Tevatron jet measurements.

Until the uncertainty in the PDFs is reduced, or a method of
analytically evaluating their uncertainties is available, the best
limits on compositeness will come from measurements insensitive to
the PDFs, and in particular, measurements that rely on angular
information. These angular distributions show the partonic hard
scattering to be well described by NLO theoretical calculations.
Comparisons of the angular and mass data to jet production models
augmented by quark compositeness show no preference for deviations
from standard QCD predictions. In fact, the analyses now show that
the compositeness scale must be greater than 2.5 TeV if it
exists at all.

The high precision Tevatron jet data has fostered a period of great
progress in QCD.  Due to the flexibility of theoretical
predictions, pQCD can describe nearly all inclusive jet and dijet
observations. Limits on quark substructure from Run 1 have nearly
doubled from the previous measurements. However, more stringent
tests will require improved PDFs and reduced theoretical
uncertainties. Looking forward, expectations are high that the
experimental measurements of jets and their properties will
continue to improve in Run 2 with a 20--fold increase in data sample
size, increased beam energy, and reduced systematic uncertainties
from the upgraded detectors.

\vspace{.2in}
\noindent ACKNOWLEDGMENTS

We wish to thank the members of the CDF and D0 Collaborations for
permitting us to include their results; in particular we would like
to thank Iain Bertrum, Anwar Bhatti, Frank Chlebana,
Bjoern Hinrichsen, Bob Hirosky and John Krane for
helpful discussions during preparation of this review and for
providing excellent illustrations.

\end{document}